\documentclass[aps
		,prb
		,twocolumn
		,superscriptaddress
		]{revtex4-2}
\usepackage[dvipsnames]{xcolor}
\usepackage{siunitx}
\usepackage{%overpic
        graphicx
        ,hyperref
        ,multirow
        ,amsmath
        ,amssymb
        ,array
        ,ulem
	    ,booktabs
        ,dcolumn
        ,pdfpages
        ,paralist
			}
\makeatletter
\AtBeginDocument{\let\LS@rot\@undefined}
\makeatother
\usepackage[version=4]{mhchem}

\newcommand{\etal}{\textit{et al.}}
\newcommand{\gao}{Ga$_2$O$_3$}
\newcommand{\betagao}{$\beta$-Ga$_2$O$_3$}
\newcommand{\Ag}{$\mathrm{A_g}$}
\newcommand{\Bg}{$\mathrm{B_g}$}

\usepackage{color}

\definecolor{darkgreen}{RGB}{0,128,0}

\begin{document}
% Hyperspectral Analysis for Complete Raman Tensor Determination in β-Ga₂O₃
\title{Complete Raman Tensor Determination in Birefringent \texorpdfstring{\betagao{}}{β-Ga₂O₃} by Single-Stage Hyperspectral Analysis of Polarization Angle-Resolved Raman Spectra }

\author{Hans Tornatzky}
\email{tornatzky@pdi-berlin.de}
\affiliation{Paul-Drude-Institut f\"ur Festk\"orperelektronik, Leibniz-Institut im Forschungsverbund Berlin e.V., Hausvogteiplatz 5–7, 10117 Berlin, Germany}

\author{Jonas Rose}
\affiliation{Paul-Drude-Institut f\"ur Festk\"orperelektronik, Leibniz-Institut im Forschungsverbund Berlin e.V., Hausvogteiplatz 5–7, 10117 Berlin, Germany}

\author{Moritz Mei\ss ner}
\affiliation{Paul-Drude-Institut f\"ur Festk\"orperelektronik, Leibniz-Institut im Forschungsverbund Berlin e.V., Hausvogteiplatz 5–7, 10117 Berlin, Germany}

\author{Benjamin M. Janzen}
\affiliation{Technische Universit\"at Berlin, Institut f\"ur Physik und Astronomie, Hardenbergstraße 36, 10623 Berlin, Germany}

\author{Zbigniew Galazka}
\affiliation{Leibniz-Institut für Kristallzüchtung, Max-Born-Str. 2, 12489 Berlin, Germany}

\author{{Juan Sebastián Reparaz}}
\affiliation{Institut de Ci\`{e}ncia de Materials de Barcelona, ICMAB-CSIC, Campus UAB, 08193 Bellaterra, Spain}

\author{Manfred Ramsteiner}
\affiliation{Paul-Drude-Institut f\"ur Festk\"orperelektronik, Leibniz-Institut im Forschungsverbund Berlin e.V., Hausvogteiplatz 5–7, 10117 Berlin, Germany}

\author{Markus R. Wagner}
\email{wagner@pdi-berlin.de}
\affiliation{Paul-Drude-Institut f\"ur Festk\"orperelektronik, Leibniz-Institut im Forschungsverbund Berlin e.V., Hausvogteiplatz 5–7, 10117 Berlin, Germany}
\affiliation{Technische Universit\"at Berlin, Institut f\"ur Physik und Astronomie, Hardenbergstraße 36, 10623 Berlin, Germany}

\date{\today}

\begin{abstract}
The low symmetry of the monoclinic phase of \gao{} leads to pronounced optical anisotropy and, consequently, to birefringence, which strongly affects the Raman response. Because Raman scattering is fundamentally sensitive to the polarizability of a material, this anisotropy must be carefully accounted for in order to extract quantitative information —-- an effort that has only recently been shown to be feasible in such media.
Here, we report Raman measurements from all three principal crystal planes (100), (010), and (001), as well as from the ($\overline{2}01$)-plane of a \betagao{} single crystal. By combining polarization angle-resolved Raman spectroscopy (PARRS) with a newly developed fitting procedure and explicitly accounting for birefringence, we achieve full spectral separation and quantitatively determine the energies and relative Raman tensor elements of all 15 Raman-active modes.
\end{abstract}

\maketitle

\section{Introduction}
The ultra-wide-band-gap semiconductor $\beta$-Ga$_2$O$_3$ has attracted significant attention as a promising material for next-generation power electronics, owing to its widely tunable electrical conductivity, high breakdown field, and the availability of large, scalable substrates for homoepitaxial growth~\cite{Pearton,Pearton2018,Higashiwaki,Higashiwaki2,Higashiwaki2017,Kokubuna2007,Oshima2008,Green2022}. 

$\beta$-Ga$_2$O$_3$ crystallizes in a low-symmetry monoclinic structure, with mutually unequal lattice parameters $a$, $b$, and $c$, and a monoclinic angle of $103.8^\circ$ between the $a$ and $c$ axes (Fig.~1a). 
This structural anisotropy gives rise to pronounced direction-dependent optical properties~\cite{Matsumoto1974,Ueda1997,Yamaguchi2004, 
Onuma2015,Sturm2015,
Varley2015,
Furthmueller2016,Onuma2016b,Ricci2016,Sturm2016b,
Mock2017,Ratnaparkhe2017,
Gopalan2020,
Cho2021,
Zhang2021optical,
Meissner2024,
Grundmann2016,Onuma2016,Schubert2016,Sturm2016,Spencer2022}, including a strongly anisotropic Raman response~\cite{
Grundmann2016,Onuma2016,Schubert2016,Sturm2016,Spencer2022,
Kranert2016PRL,Zhang2022,VonWenckstern2017,
Parisini2018,
Ghosh2016,Villora2002,
Look2019,Ritter2019,Mengle2019,Janzen2022,
Janzen2021_beta,Dohy1982,Kranert2016SciRep,Liu2007}. 
Owing to its biaxial optical symmetry~\cite{Grundmann2017}, the dielectric tensor of $\beta$-Ga$_2$O$_3$ exhibits three distinct diagonal elements, 
$\varepsilon_{\tilde{x}\tilde{x}}$, 
$\varepsilon_{\tilde{y}\tilde{y}}$, and 
$\varepsilon_{\tilde{z}\tilde{z}}$, when expressed in its eigensystem 
$(\tilde{x}$, $\tilde{y}$, $\tilde{z})$~{\cite{Kai-thermal}}.
As a consequence, birefringence arises for most light-propagation directions, posing a challenge for the quantitative analysis of Raman scattering --- an effort long regarded as “pointless’’ in optically anisotropic media~\cite{Beattie1968}.
For this reason, most Raman studies in the literature have neglected birefringence and instead focused on properties that remain unaffected in polarised~\cite{Machon2006,Onuma2014,Yao2019,Zhang2021_beta,Janzen2021_beta,Janzen2022,Seyidov2022} or temperature-dependent measurements~\cite{Dohy1982,Fiedler2020,Zhang2021_T-dpendent}. 
Only few studies have explicitly accounted for birefringence: either by introducing a complex Raman tensor~\cite{Zhang2022}, or by incorporating the birefringence-dependent physical processes that accompany Raman scattering directly into the analytical model~\cite{Kranert2016SciRep}.

In this work, we use polarized, angle-resolved Raman scattering (PARRS) adapting the model proposed by Kranert \etal{}~\cite{Kranert2016PRL, Kranert2016SciRep}, to calculate the PARRS profiles, i.\,e.\ the polarization angle dependent intensity, in birefringent media.  
We develop a new fitting procedure for the hyperspectral data which improves accuracy and enables reliable differentiation of strongly overlapping phonon modes of different symmetries. Applied to  \betagao{}, this procedure enables the reliable separation of all Raman active modes and, hence, to  determine their energies and relative Raman tensor elements, including the previously unresolved mode pairs $\mathrm{A_{g}^{(5)}/B_{g}^{(3)}}$, $\mathrm{A_{g}^{(7)}/B_{g}^{(4)}}$ and $\mathrm{B_{g}^{(5)}/A_{g}^{(9)}}$.

%###############################
%###############################
%###############################
\section{Experimental Methods}
Raman scattering measurements were performed at room temperature using the 633\,nm line of a HeNe laser and a custom-modified LabRAM HR 800 spectrometer (Horiba Jobin-Yvon). The laser was focused onto the sample through a 50$\times$ Olympus microscope objective (NA = 0.55), and scattered light was collected in backscattering geometry. Samples were positioned with their surface normal along the light propagation direction.
Inelastically scattered light was dispersed by the 800\,mm monochromator equipped with a 1800 \,lines/mm grating and detected by a CCD.  A half-wave plate ($\lambda/2$) in the excitation path was set at 0° or 45° to align the incident linear polarization parallel or perpendicular to the detected scattered light, which was selected with a fixed polarizer. 
Raman spectra were calibrated using multiple neon emission lines. 
Polarization angle-resolved Raman measurements (PARRS) were performed by introducing an additional half-wave plate between notch filter and the objective, which rotates the polarization of both the incident ($\hat{e}_{\mathrm{i}}$) and scattered ($\hat{e}_\mathrm{s}$) light by an angle $\phi$ relative to the fixed sample orientation, effectively simulating a sample rotation. The wave plate is mounted in a motorized, computer-controlled holder, enabling automated recording of polarization-angle-dependent intensity profiles to determine relative Raman tensor elements~\cite{Kranert2016SciRep,Kosc2020,Sander2012} or the thickness of thin films~\cite{Hildebrandt2021}.

Raman spectra were acquired for (100)-, (010)-, (001)- and ($\overline{2}01$)-oriented \betagao{} single crystals. The samples were cut from the same undoped, two-inch-diameter crystals grown by the Czochralski method at Leibniz-Institut für Kristallzüchtung using an Ir crucible and an oxidizing growth atmosphere~\cite{Galazka2017,Galazka2021}, and chemically–mechanically polished. Hall-effect measurements indicate a free-electron concentration of $3.4 \times 10^{17}$~cm$^{-3}$ and an electron mobility of 118~cm$^2$V$^{-1}$s$^{-1}$~\cite{Galazka2022b}. %\cite{Irmscher2011}
The crystals exhibit high structural quality, with no twins, low-angle grain boundaries, or nanopipes, and a dislocation density below $\sim 10^3$~cm$^{-3}$~\cite{Galazka2022}.

A newly developed hyperspectral fitting procedure was used for the analysis and is described below. Spectroscopic analysis was performed in part using a non-public alpha version of \textit{peak-o-mat2}; the stable version is publicly available on GitHub~\cite{pom-git,qceha}.

%###############################
%###############################
%###############################
\section{Results and discussion}

%############################### Fig 1 %###############################
\begin{figure}[tbh]
\centering
a) \hfill \phantom{.}\\
\vspace{-4ex}
\includegraphics[trim = 145 0 00 0, clip, width=.48\textwidth]{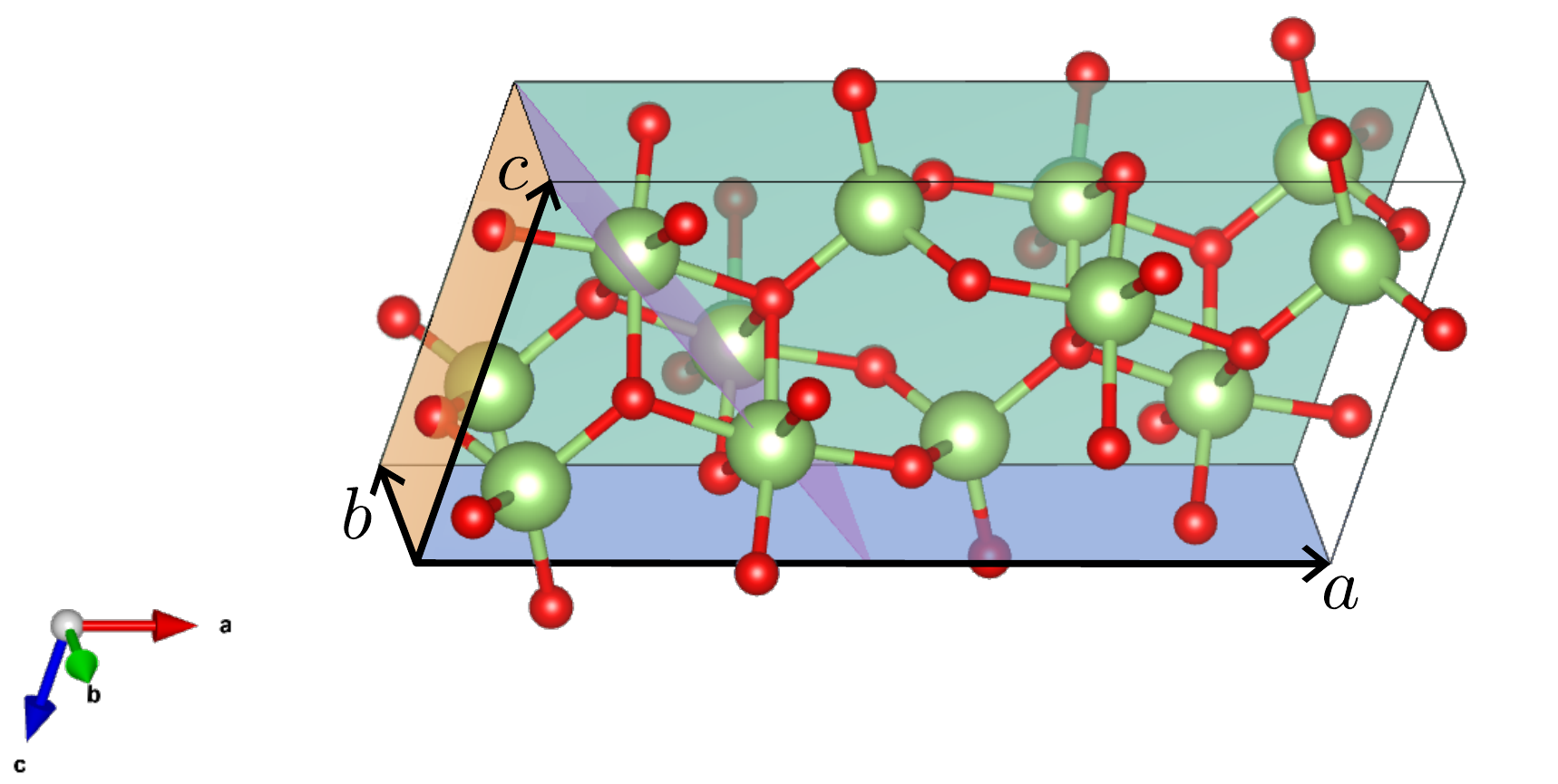}
\vspace{0ex}
 b) \hfill \phantom{.}\\
\vspace{-10ex}
\includegraphics[width=0.40\textwidth]{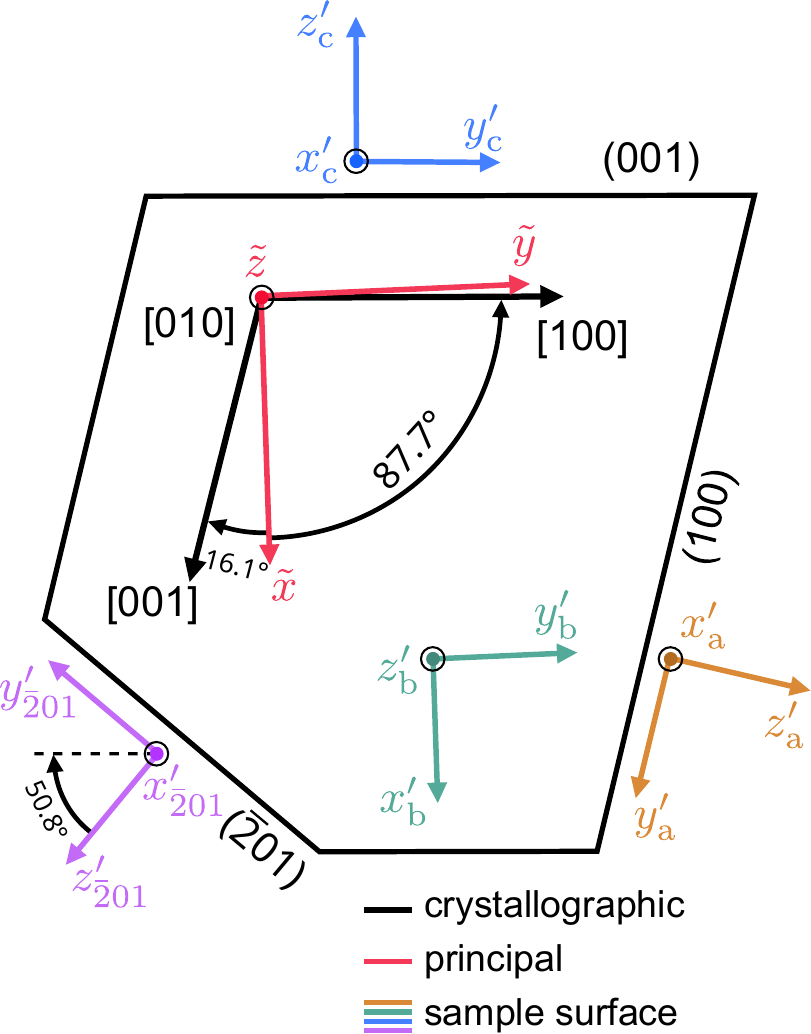}
\caption{a) Unit cell of \betagao{}.
The (100), (010), (001) and ($\overline{2}$01) planes are indicated in different colors. The crystallographic plot was created using VESTA~\cite{Momma2011}. b) Relation between the crystallographic, principal, and sample surface coordinate systems in \betagao.
}
\label{fig:CS}
\end{figure}
%############################### Fig 1 %###############################

The monoclinic unit cell of \betagao{} contains 10 atoms, giving rise to 30 phonon branches. At the $\Gamma$ point they decompose into the irreducible representations of the $\mathrm{C_{2h}}$ point group \cite{Kroumova}
\begin{align*}
\Gamma &= 5 \mathrm{A_u} + 10 \mathrm{B_u} + 10 \mathrm{A_g} + 5\mathrm{B_g},
\end{align*}
where \mbox{$\Gamma_{\mathrm{aco}}=\mathrm{A_u} + 2 \mathrm{B_u}$} are the acoustic phonons.
The optical modes of even parity (index $\mathrm{g}$) are Raman-active with Raman tensors of the form
\begin{equation}
\setlength{\jot}{18pt}
  \mathcal{R}: \mathrm{A_g}= \left(\begin{matrix}a&d&0\\d&b&0\\0&0&c\end{matrix}\right),\  \mathrm{B_g}=\left(\begin{matrix}0&0&e\\0&0&f\\e&f&0\end{matrix}\right),
\label{eq:Ramantensors}
\end{equation}
whereas the modes of odd parity (index u) are IR-active. 

We introduce three different coordinate systems, as depicted in figure~\ref{fig:CS}b), to facilitate the discussion of the crystal's properties and the analysis: 
(1) The crystallographic coordinate system (black) where $a$, $b$, and $c$ are aligned with the crystallographic directions [100], [010], and [001], respectively. $a$ and $c$ enclose the monoclinic angle $\beta = 103.8^\circ$. 
(2) The orthonormal principal axis coordinate system (red), where $\tilde{x}$, $\tilde{y}$, and $\tilde{z}$ are defined by the eigenvectors of the dielectric tensor, where $\tilde{z}\parallel [010]$ due to their natural coincidence. We note that the $\tilde{x}$ and $\tilde{y}$ vectors' orientation within the (010)-plane and absolute magnitude are wavelength dependent. For an excitation wavelength of \SI{632.8}{\nano\meter}, $\tilde{y}$ and $[100]$ enclose an angle of $\alpha=2.33^\circ$ and the dielectric tensor reads in its eigensystem \cite{Sturm2015,Sturm-Mail} 
\begin{equation}
    \varepsilon = \begin{pmatrix}
        3.727 & 0 & 0 \\
        0 & 3.624 & 0 \\
        0 & 0 & 3.768
    \end{pmatrix}.
\label{eq:dielectrict}
\end{equation}
This is the coordinate system in which the Raman tensors in equation~\ref{eq:Ramantensors} are defined.
(3) The orthonormal sample coordinate systems ($x'_p$, $y'_p$, $z'_p$) for each of the investigated crystal planes $p$. $z'_p$ is chosen to align to the surface normal, i.\,e.\ to coincide with the microscope's optical axis. Consequently, $x'_p$ lies within the $p$-plane and is chosen to be aligned with a principal axis; on the $b$-plane $x'_b \parallel \tilde{x}$, whereas $x'_p \parallel \tilde{z} \parallel [010]$ on the other investigated planes.
We note that another commonly used system is the orthonormal quasi-crystallographic coordinate system where two out of ($x$, $y$, $z$) are chosen to be parallel to the $a$ and $b$ lattice vectors. Different definitions are common to different fields, so care should be taken when comparing anisotropic properties~\cite{Kai-thermal}. However, this system is not required to understand the arguments in this study.
A detailed description of the above systems and their respective transformations is given in the supplemental material~\cite{supplement}.\\

To calculate the Raman selection rules for a given surface geometry, one typically uses the well-established expression
\begin{equation}
I\propto \left|{\hat{e}_{\mathrm{s}}} {R} {\hat{e}_{\mathrm{i}}} \right|^2,
\end{equation}
where $R$ is the Raman tensor and $\hat{e}_{\mathrm{i}}$ and $\hat{e}_{\mathrm{s}}$ are the incident and scattered polarization vectors, respectively, commonly transformed from the dielectric tensor system ($\tilde{x}$, $\tilde{y}$, $\tilde{z}$) (cf. Eq.\ \ref{eq:Ramantensors}) into the respective surface system ($x'_p$, $y'_p$, $z'_p$). For parallel scattering configuration, they are given by $\hat{e}^{\parallel}_{\mathrm{i}} = \hat{e}^{\parallel}_{\mathrm{s}} = (\cos(\varphi), \sin(\varphi), 0)$, whereas for crossed configuration they read $\hat{e}^{\perp}_{\mathrm{i}} = (\cos(\varphi), \sin(\varphi), 0)$ and $\hat{e}^{\perp}_{\mathrm{s}} = (-\sin(\varphi), \cos(\varphi), 0)$. Here, $\varphi$ denotes the surface system polarization angle, which is related to the measurement polarization angle $\phi$ through $\phi = \varphi - \varphi_0$. Hence, $\varphi_0$ denotes the offset between the crystal orientation with respect to the setup-defined origin of the polarization, so that
$\varphi=0^\circ$ corresponds to $\hat{e}_{\mathrm{i}}\parallel x'_p$ and $\varphi=90^\circ$ to $\hat{e}_{\mathrm{i}}\parallel y'_p$. \\
Here, for \betagao{}, the influence of birefringence must be taken into account when calculating the angle-dependent selection rules. To this end, we follow the approach proposed by Kranert \textit{et al.}~\cite{Kranert2016SciRep,Kranert2016PRL}, which is briefly summarized in the following.\\

The most fundamental consequence is that the polarization vectors of the incident and scattered radiation cannot be regarded as fixed during propagation through the birefringent crystal because their relative phase is constantly retarded. However, for excitations within the transparency regime, i.\,e.\ large penetration depths, as well as pronounced birefringence, this effect can be accounted for by employing
\begin{equation}
    I \propto \left| \hat{e}_{\rm s} R_0 \hat{e}_{\rm i} \right|^2 + \left| \hat{e}_{\rm s} R_1 \hat{e}_{\rm i} \right|^2 + \left| \hat{e}_{\rm s} R_2 \hat{e}_{\rm i} \right|^2,
\label{eq:intensity_biref}
\end{equation}
where
\begin{equation}
    R_0 = \begin{pmatrix} r_0 & 0\\0 & 0 \end{pmatrix}, \quad R_1=\begin{pmatrix} 0 & r_1\\r_1 & 0 \end{pmatrix}, \quad R_2=\begin{pmatrix} 0 & 0\\0 & r_2 \end{pmatrix}
\label{eq:Ramanmat}
\end{equation}
are summands of the Raman tensor represented in the corresponding two-dimensional surface system ($x'_p$, $y'_p$), such that $R=R_0+R_1+R_2$.\\
Another consequence of optical anisotropy is that the eigenpolarizations inside the crystal may exhibit an out-of-plane component when the propagation direction does not coincide with one of the principal axes. Consequently, the external polarization vectors are transformed into the allowed internal polarizations of the crystal by using the transformation matrix
\begin{equation}
    T = 
    \begin{pmatrix}
    1 & 0\\
    0 & \varepsilon_{z'z'} \left(\varepsilon^2_{y'z'}+\varepsilon^2_{z'z'}\right)^{-1/2}\\
    0 & \varepsilon_{y'z'} \left(\varepsilon^2_{y'z'}+\varepsilon^2_{z'z'}\right)^{-1/2}
    \end{pmatrix}.
\end{equation}
Here, $\varepsilon_{y'z'}$ and $\varepsilon_{z'z'}$ denote the corresponding components of the dielectric tensor transformed into the surface system in the same way as the Raman tensor. In addition, a transmission correction is introduced which takes into account the different Fresnel transmission coefficients of s- and p-polarized light at the crystal surface. This correction is implemented by applying the diagonal matrix $\rho = \mathrm{diag}(t_{x'x'}/t_{y'y'},1)$ to both the incident and scattered polarization vectors. Here, $t_{x'x'}=2/(\sqrt{\varepsilon_{x'x'}}+1)$ denotes the Fresnel transmission coefficient for normal incidence along $z'_p$ with polarization parallel to $x'_p$, while $t_{y'y'}$ is defined analogously for polarization along $y'_p$. Similarly, birefringence affects the effective solid angle from which scattered light is collected by the microscope objective
by changing the refraction of the collection cone inside the crystal. To account for this effect, the solid angle $\Omega = 4\pi\sin^2\left(\arcsin\left(NA/\sqrt{\varepsilon}\right)/2\right)$ is evaluated for different refractive indices and the correction matrix $Z = \mathrm{diag}(\sqrt{\Omega_{x'x'}/\Omega_{y'y'}}, 1, 1)$ is applied to the transformed Raman tensor.\\
In conclusion, the Raman selection rules are obtained by first constructing the effective transformed Raman tensor
\begin{equation}
    R_{\mathrm{eff}} = \rho T^\intercal ZRT\rho
\label{eq:efframan}
\end{equation}
and subsequently inserting it into equation \ref{eq:intensity_biref}, using the decomposition of the Raman tensor defined in equation~\ref{eq:Ramanmat}. 

Finally, to derive the numerical values of the tensor elements, one has to account for a wavelength dependent prefactor that originates in the nature of dipole scattering~\cite{HayesWilliam1978Solb}:
\begin{equation}
    C(\omega_{\mathrm{p}}) = \frac{\omega_{\mathrm{i}}(\omega_{\mathrm{i}}-\omega_{\mathrm{p}})^{3}}{\omega_\mathrm{p}\left[1-\mathrm{exp}\left(-\frac{\hbar\omega_{\mathrm{p}}}{\mathrm{k_{B}}T}\right)\right]}.
\end{equation}

\begin{figure}[htb]
\includegraphics[clip, width=\linewidth]{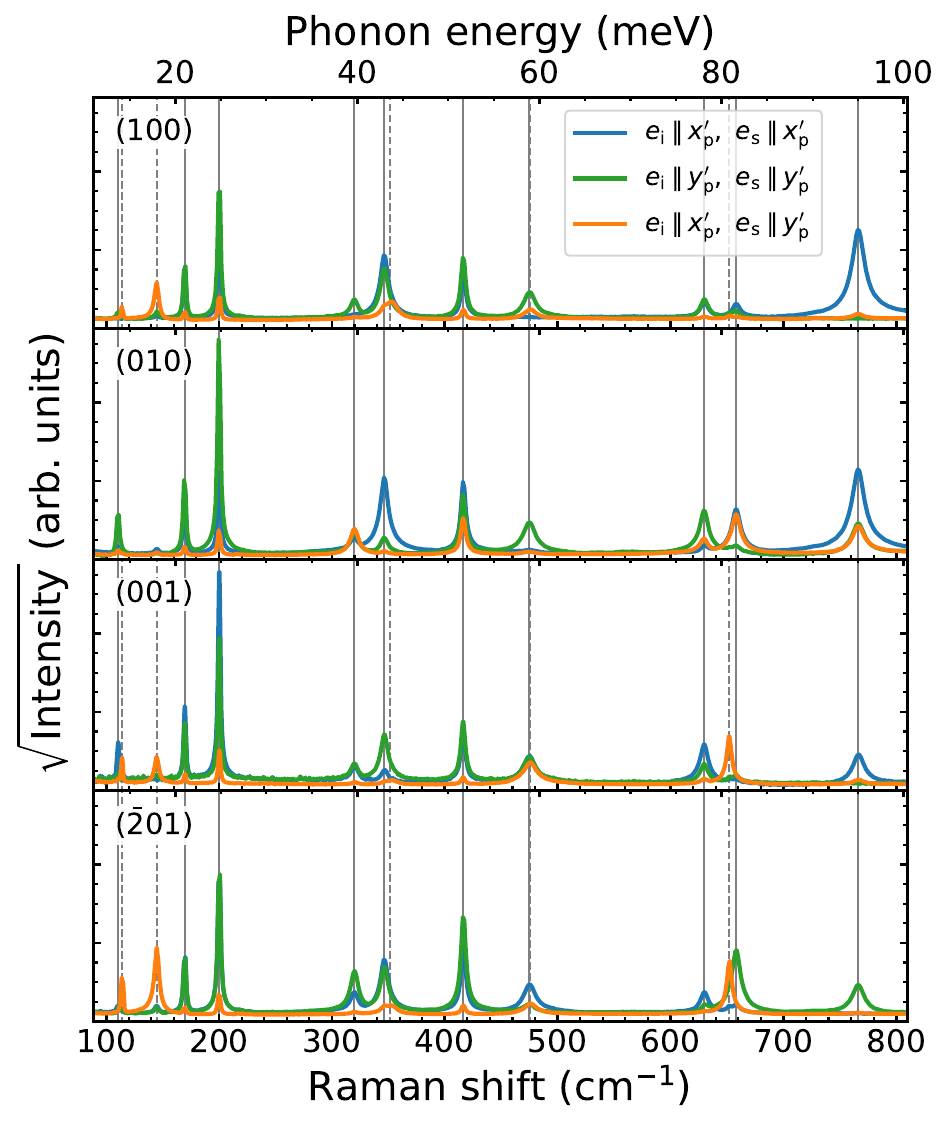}
\caption{Polarized Raman spectra as measured on the four investigated crystal planes in parallel and crossed polarization scattering geometries. Due to the two-fold rotational symmetry in the parallel polarized spectra, both spectra, 
with the polarization aligned along one of the sample surface axes ($x'_p$, $y'_p$) are plotted. 
The Raman spectra are depicted with the square root of the intensity for better visualization of small peaks.}
\label{fig:Raman_spectra}
\end{figure}

Figure~\ref{fig:Raman_spectra} displays Raman spectra resolving all 15 Raman-active phonons \mbox{($10 \mathrm{A_g} + 5 \mathrm{B_g}$)} via a set of dedicated polarization geometries. Scattering geometries are indicated using the Porto notation 
$\hat{k}_\mathrm{i}(\hat{e}_\mathrm{i}\hat{e}_\mathrm{s})\hat{k}_\mathrm{s}$, 
where $\hat{k}_\mathrm{i}$ and $\hat{k}_\mathrm{s}$ denote the propagation directions of the incident and scattering light, respectively.
The combination of polarization-dependent measurements and Raman selection rules allows the selected detection and differentiation of all $\mathrm{A_g}$ and $\mathrm{B_g}$ modes. 
{This enables the determination of the phonon frequencies for the closely-spaced $\mathrm{A_{g}^{(5)}/B_{g}^{(3)}}$, $\mathrm{A_{g}^{(7)}/B_{g}^{(4)}}$ and $\mathrm{B_{g}^{(5)}/A_{g}^{(9)}}$ mode pairs despite their spectral overlap from the respective other mode. }

%############################### Fig 2 %###############################
%############################### end Fig 2 %###############################

As predicted by Raman selection rules, the \Ag{} modes experience (local) maximum intensity in parallel-polarization scattering geometries with $\hat{e}_\mathrm{i}$ and $\hat{e}_\mathrm{s}$ parallel to one of the principal directions (i.\,e.\ $\tilde{x}$, $\tilde{y}$ or $\tilde{z}$), while modes of \Bg{} symmetries are absent in these scattering conditions.
To acquire spectra only allowing \Bg{} modes, (100), (001) and ($\overline{2}01$) planes are measured with crossed polarization, while $\hat{e}_\mathrm{i}$ and $\hat{e}_\mathrm{s}$ are parallel to different principal directions.
When exciting along the [010] direction, $\mathrm{B_{g}}$ modes are always prohibited according to Raman selection rules, regardless of the selected polarization configuration. 
{By fitting Lorentzian line shape functions to the spectra depicted in figure~\ref{fig:Raman_spectra}, we determine the peak positions of all 15 Raman-active phonon modes (listed in \mbox{Tab. \ref{tab:tensorelements_all}}). The obtained results show good agreement with previous experimental~\cite{Kranert2016SciRep,Janzen2021_beta,Dohy1982,Machon2006,Onuma2014,Janzen2022} and theoretical~\cite{Kranert2016SciRep,Janzen2021_beta,Dohy1982,Machon2006,Liu2007} studies.}

%############################### Fig 3 %###############################
\begin{figure}[tbh]
\includegraphics[width=\linewidth]{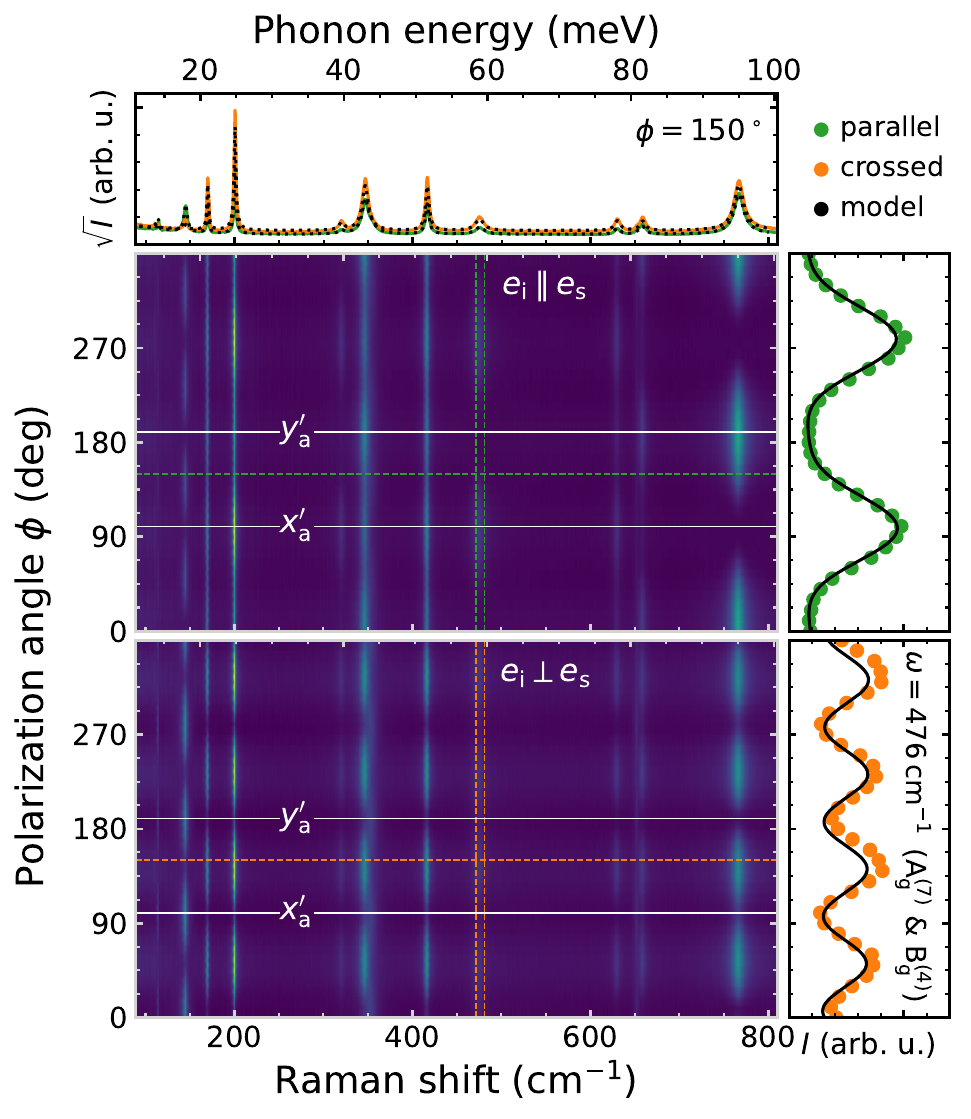}
\caption{Contour plot of the polarization dependent Raman spectra on the a-plane for parallel (middle) and crossed scattering geometries (bottom). Angles, where $x'_{\mathrm{a}} \parallel [010]$ and $y'_{\mathrm{a}} \parallel [001]$ are labeled. In the top panel Raman spectra with a polarization angle of {$\phi=150^\circ$} are depicted, while the panels on the right show the integrated intensity in the range of the $\mathrm{A^{(7)}_g}$/$\mathrm{B^{(4)}_g}$ pair and the corresponding slice of the fit. The Raman spectra and contour plots are depicted with the square root of the intensity for better visualization of small peaks.  
% "These offset angles are: 81.26° (a), 75.95° (b), 1.86° (c) and 96.34° 
}
\label{fig:PARRS}
\end{figure}
%############################### end Fig 3 %###############################

We extend this investigation by removing the constraint that the light polarization must be parallel to the principal axes and instead measure Raman spectra in parallel and crossed configuration 
rotating the light polarization around the crystal surface normal.
Figure~\ref{fig:PARRS} depicts a representative contour plot of these PARRS spectra on the a-plane in parallel (top) and crossed  polarization (bottom) for a sequence of rotational polarization angles $\phi$ with respect to the fixed sample. $\phi = 0$ marks the onset of the measurement. Angles coincident with the principal axes $x'_\mathrm{a} \parallel [010]$ and $y'_{\mathrm{a}} \parallel [001]$ are labeled and correspond to the spectra plotted in figure \ref{fig:Raman_spectra}. 
For quantitative analysis, PARRS profiles, i.\,e.\ the Raman intensity of each mode expressed as a function of the rotational angle $\phi$, are conventionally obtained by individually fitting the Raman spectra and plotting the peak's intensities over the polarization angle $\phi$ to fit the relative Raman tensor elements in a consecutive step. Here, we present a new approach where the hyperspectral data is fitted in a single step to obtain the phonon energies, full width at half maximum, and the relative tensor elements. 
The fit function mainly comprises a two-dimensional function for each of the scattering geometries (sg), i.\,e.\ for all combinations of surface plane and polarization configuration:
\begin{equation*}
    F_\text{sg}(\omega, \varphi) = s_\text{sg}\sum_i L_i(\omega) A_{i,\text{sg}}(\varphi)+B_\text{sg}(\omega, \varphi),
\end{equation*}
where $s_\text{sg}\approx 1$ is a scaling factor to account for slight differences in focus and surface roughness induced reflectances, and 
$L_i$, $A_i$ and $B_\text{sg}$ 
denote the Lorentzian lineshape and the PARRS profile of the i-th Raman-active mode, and the background function, respectively.
The $L_i$ further include the phonon energy and line broadening, while the $A_i$ contain parameters of the crystal's symmetry and scattering properties, i.\,e.\ the relative Raman tensor elements. Additional parameters are included to account for experimental artifacts such as a background, slight differences in focus, etc. 

% In the presented study the hyperspectral data comprises of 296 spectra (37 spectra on 4 crystallographic planes for parallel and crossed polarization).

Although the presented fitting procedure shows negligible benefits for materials yielding Raman spectra with well isolated peaks, the opposite is true when peaks strongly overlap. By including the modes' predicted angle responses, it becomes possible to reliably fit spectra with peaks overlapping beyond the spectral resolution, where fits of a single spectrum yield peak intensities with extensive errors. Therefore, in \betagao{}, the newly presented procedure yields a  significantly better fit to the hyperspectral data set.

Slices of the fit result are presented alongside the exemplary Raman spectrum and angle dependence of the $\mathrm{A^{(8)}_g}$ 
mode in the top and right panels in figure~\ref{fig:PARRS} as a solid black line. The low deviation from the experimental data provides evidence of an excellent fit. The complete set of data including all PARRS contour plots and plots of the fitted model and residuals is provided in the supplemental material~\cite{supplement}.

It is worth mentioning that the determined Raman tensor elements are relative quantities and are therefore typically scaled in such a way that the largest tensor element has a value of 1000; here, this is the case for the $a$ parameter of the $\mathrm{A_{g}^{(10)}}$ mode. 
The normalized relative Raman tensor elements are listed in table~\ref{tab:tensorelements_all}. Our determined tensor elements agree well with most of the experimental values reported in Ref.~\cite{Kranert2016SciRep}. Some deviations are likely related to systematic differences, including variations in the experimental setup
%, i.\,e. the optical components, 
and the fitting procedure.
% In Ref.\ \cite{Zhang2022}, it was demonstrated that different excitation wavelengths in the transmission regime of \betagao{} can also lead to variations in the relative Raman intensities. 
However, the individual contribution of the systematic errors %and the wavelength-dependency of the Raman tensor elements 
cannot be distinguished based on the present comparison. Yet, we consider the major contribution to originate from an improved accuracy obtained by the hyperspectral fitting, as this approach enables the determination of the Raman tensor elements for the $\mathrm{B^{(3)}_g}$ and $\mathrm{B^{(4)}_g}$ modes, which could previously not be determined experimentally and which are in good agreement with the values predicted by theory \cite{Kranert2016SciRep}.
% \newline

Since Raman spectra are measured intensities, i.\,e.\ formally a quadratic function of the tensor elements, it is often considered that the tensor elements' signs cannot be determined. This, however, is not necessarily true when scattering information from multiple planes is available: the functions describing the PARRS profiles on planes where the surface normal is not a principal axis of the dielectric tensor will include linear combinations the tensor elements.
These combinations can reveal from the experimental data whether their signs are equal or opposite.
Here, due to the set of investigated planes, the tensor elements group as ($a$, $b$, $d$) and ($e$, $f$). Hence, fixing the sign of $a$ and $e$ to be positive yields the \mbox{relative} sign of $b$, $d$ and $f$, respectively. Solely the element $c$ enters quadratically into all fit formulae (cf.~Tab.~\ref{tab:selectionrules} in the SI). 
Crystal planes excluding those that contain $\tilde{z}$ within the surface and the (010) plane, such as the (110)-plane, link the $\mathrm{A_g}$-modes' $c$ element to the $(a, b, d)$ group, thereby enabling the determination of the sign of $c$ relative to $a$.

\begin{table}%[htb!]
\caption{Phonon mode irreducible representations, determined frequencies $\omega_{\mathrm{p}}$, Raman tensor elements $a$, $b$, $c$, $d$, $e$, $f$, and scattering strength prefactors $C(\omega_{\mathrm{p}})$ of the Raman active phonons as obtained by polarization angle-resolved Raman measurements. 
The tensor elements and $C(\omega_{\mathrm{p}})$ prefactors
are normalized to a value of $a=1000$ for the  $\mathrm{A_{g}^{(10)}}$ mode and to $C(\mathrm{A_{g}^{(1)}})=1$. 
The error of the phonon energies can be considered  well below the spectral resolution of 0.5\,cm$^{-1}$.}
\label{tab:tensorelements_all}
\begin{tabular}{ccccccccc}
\toprule \toprule
% \toprule[1pt]
\multicolumn{1}{c}{}  & $\omega_{\mathrm{p}}$ & $a$ & $b$ & $|c|$ & $d$ & $e$ & $f$ & $C(\omega_{\mathrm{p}})$\\
\multicolumn{1}{c}{}  & ($\mathrm{cm^{-1}}$) & & & & & & &\\
\midrule[0.5pt]
    $\mathrm{A^{(1)}_g}$ & 111.4 & 18 & $-$55 & 18 & $-$2 & & & 1.000\\
    $\mathrm{A^{(2)}_g}$ & 170.4 & 102 & 140 & 122 & $-$7 & & & 0.479\\ 
    $\mathrm{A^{(3)}_g}$ & 200.9 & 186 & 429 & 321 & 11 & & & 0.364\\
    $\mathrm{A^{(4)}_g}$ & 320.9 & 96 & 147 & 143 & 125 & & & 0.176\\ 
    $\mathrm{A^{(5)}_g}$ & 347.2 & 428 & 123 & 324 & $-$13 & & & 0.157\\
    $\mathrm{A^{(6)}_g}$ & 417.0 & 370 & 301 & 346 & 147 & & & 0.121\\
    $\mathrm{A^{(7)}_g}$ & 475.6 & 64 & $-$300 & 306 & $-$35 & & & 0.101\\
    $\mathrm{A^{(8)}_g}$ & 630.4 & 56 & 383 & 232 & $-$154 & & & 0.070\\
    $\mathrm{A^{(9)}_g}$ & 658.5 & 433 & 97 & 134 & 317 & & & 0.066\\
    $\mathrm{A^{(10)}_g}$ & 766.9 & 1000 & 382 & 96 & $-$323 & & & 0.055\\[1ex]
    % \hline
    $\mathrm{B^{(1)}_g}$ & 114.8 & & & & & 34 & 34 & 0.946\\
    $\mathrm{B^{(2)}_g}$ & 145.5 & & & & & 119 & 71 & 0.626\\ 
    $\mathrm{B^{(3)}_g}$ & 353.5  & & & & & {169} & {$-$37} & 0.153\\
    $\mathrm{B^{(4)}_g}$ & 476.6 & & & & & {35} & {$-$236} & 0.101\\
    $\mathrm{B^{(5)}_g}$ & 652.6 & & & & & 148 & 359 & 0.067\\
\bottomrule \bottomrule
% \bottomrule[1pt]
\end{tabular}
\end{table}

\section{Conclusions}
We have applied polarization angle-resolved Raman spectroscopy (PARRS) to determine the Raman tensor elements of the Raman-active phonons of \betagao{} within an extended tensor formalism that accounts for optical anisotropy and birefringence. By combining measurements on multiple crystal planes with a hyperspectral single-stage fitting procedure, we obtain the phonon energies and relative Raman tensor elements of all 15 Raman-active modes in a self-consistent manner.

This approach enables reliable analysis of strongly overlapping Raman features and yields tensor elements for modes that could not be determined in earlier studies, including the $\mathrm{B^{(3)}_g}$ and $\mathrm{B^{(4)}_g}$ modes. {These results show that a procedure to model the hyperspectral data in a single step is essential for a quantitative analysis of polarized Raman scattering in monoclinic \betagao{} }and provide a complete experimental benchmark for comparison with theory and for future studies of optically anisotropic materials.

% Polarization angle-resolved Raman spectroscopy {(PARRS)} was applied to determine the Raman tensor elements of the individual Raman-active phonon modes of \betagao{} in the framework of an extended Raman tensor formalism for optically anisotropic materials. 
% We propose a new scheme to fit the data in a hyperspectral (i.\,e. multidimensional) single-stage fit, which provides access to modes whose signal is obscured by other peaks.
% We determine all Raman-active mode energies and relative raman tensor elements, including those of the $B^{(3)}_g$ and $B^{(4)}_g$ modes that could not be determined in previous studies. 

\begin{acknowledgments}
Chris Sturm (Universität Leipzig) and Christian Kranert ({Fraunhofer Technology Centre Semiconductor Materials THM}) are gratefully acknowledged for helpful discussions and providing the dielectric tensor elements of $\beta$-Ga$_2$O$_3$ for \mbox{632.8 nm}. We thank Christian Kristukat for support with the software analysis of the polarization angle-resolved Raman spectra. Andreas Fiedler (Leibniz-Institut für Kristallzüchtung) is acknowledged for critical reading of the paper.
Funding by the BMFTR, Berlin Senate and Deutsche Forschungsgemeinschaft (DFG, German Research Foundation) under project number 446185170, by the Bundesministerium für Bildung und Forschung (BMBF) project under Grant No.\ 16ES1084K, by the MCIN/AEI/10.13039/501100011033 under grant PID2024-162811NB-I00 and the Severo Ochoa Centres of Excellence Program under grant CEX2023-001263-S are acknowledged.
\end{acknowledgments}

\bibliographystyle{apsrev4-2}
\section*{References}

\bibliography{bibliography-betagao}

@misc{supplement,
    note= {See Supplemental Material at [provided by Publisher] for a detailed description of the coordinate systems and their respective transformations as well as the comprehensive set of contour plots, plots of the fitted model function and respective fit residual.}
}

@book{HayesWilliam1978Solb,
    series = {A Wiley-Interscience publication},
    publisher = {Wiley},
    isbn = {0471031917},
    year = {1978},
    title = {Scattering of light by crystals},
    Tlanguage = {eng},
    address = {New York},
    author = {Hayes, William and Loudon, Rodney},
    pages = {31}
}

@unpublished{Kai-thermal,
    author = {Kai Xu and Shuo Zhao and Luca Sung-Min Choi and Nils Bernhardt and Pouria Emtenani and Moritz Meißner and Hans Tornatzky and Oliver Bierwagen and Piero Mazzolini and Riccardo Mincigrucci and Laura Foglia and Danny Fainozzi and Filippo Bencivenga and Zbigniew Galazka and George Fytas and Bartlomiej Graczykowski and Riccardo Rurali and Matthias Scheffler and Christian Carbogno and Juan Sebastian Reparaz and Markus R. Wagner},
    title = {{Nanoscale origin of thermal anisotropy in monoclinic $\beta$-Ga$_2$O$_3$}},
    note = {{Nanoscale origin of thermal anisotropy in monoclinic $\beta$-Ga$_2$O$_3$}, in preparation}
}

@article{Meissner2024,
    author = {Meißner, Moritz and Bernhardt, Nils and Nippert, Felix and Janzen, Benjamin M. and Galazka, Zbigniew and Wagner, Markus R.},
    title = {Anisotropy of optical transitions in $\beta$-Ga$_2$O$_3$ investigated by polarized photoluminescence excitation spectroscopy},
    journal = {Appl. Phys. Lett.},
    volume = {124},
    number = {15},
    pages = {152102},
    year = {2024},
    month = {04},
    Xdoi = {10.1063/5.0189751},
    url = {https://Xdoi.org/10.1063/5.0189751},
}

@article{Galazka2022b,
    author = {Galazka, Zbigniew and Ganschow, Steffen and Seyidov, Palvan and Irmscher, Klaus and Pietsch, Mike and Chou, Ta-Shun and Bin Anooz, Saud and Grueneberg, Raimund and Popp, Andreas and Dittmar, Andrea and Kwasniewski, Albert and Suendermann, Manuela and Klimm, Detlef and Straubinger, Thomas and Schroeder, Thomas and Bickermann, Matthias},
    title = {Two inch diameter, highly conducting bulk $\beta$-Ga$_2$O$_3$ single crystals grown by the Czochralski method},
    journal = {Appl. Phys. Lett.},
    volume = {120},
    number = {15},
    pages = {152101},
    year = {2022},
    month = {04},
    issn = {0003-6951},
    Xdoi = {10.1063/5.0086996},
    url = {https://Xdoi.org/10.1063/5.0086996},
}

@article{Kroumova,
author = {E. Kroumova and M.I. Aroyo and J.M. Perez-Mato and A. Kirov and C. Capillas and S. Ivantchev and H. Wondratschek},
title = {Bilbao Crystallographic Server : Useful Databases and Tools for Phase-Transition Studies},
journal = {Ph. Transit.},
volume = {76},
number = {1-2},
pages = {155--170},
year = {2003},
publisher = {Taylor \& Francis},
Xdoi = {10.1080/0141159031000076110},
URL = {https://Xdoi.org/10.1080/0141159031000076110},
Xeprint = {https://Xdoi.org/10.1080/0141159031000076110}
}

@misc{Sturm-Mail,
author = {{C. Sturm (Universität Leipzig)}},
note = {{Private communication.}},
}

@misc{pom-git,
  author = {Kristokat, Chr.},
  title = {peak-o-mat2},
  year = {2026},
  publisher = {GitHub},
  journal = {GitHub repository},
  howpublished = {\url{https://github.com/kurisutsukato/peak-o-mat.git}},
  commit = {1113ca03e8710a7e530fd6674891fd6fecde20b4},
  note = {Analysis was performed in part with a non-public alpha-version. }
}

@misc{qceha,
  author = {Kristokat, Chr.},
  title = {peak-o-mat},
  year = {2026},
  note = {\url{https://qceha.net/}},
}

@article{Beattie1968,
Xdoi = {10.1098/rspa.1968.0199},
issn = {0080-4630},
journal = {Proc. R. Soc. Lond. A. Math. Phys. Sci.},
month = {nov},
number = {1491},
pages = {407--429},
author = {Beattie, I. R. and Gilson, T. R. },
publisher = {The Royal Society London},
title = {{Single crystal laser Raman spectroscopy}},
url = {https://royalsocietypublishing.org/Xdoi/epdf/10.1098/rspa.1968.0199 https://royalsocietypublishing.org/Xdoi/10.1098/rspa.1968.0199},
volume = {307},
year = {1968},
}

@article{Cho2021,
author = {Cho, Jeong Bin and Jung, Gunwoo and Kim, Kyuheon and Kim, Jihun and Hong, Soon Ku and Song, Jung Hoon and Jang, Joon Ik},
Xdoi = {10.1021/acs.jpcc.0c08413},
issn = {19327455},
journal = {J. Phys. Chem. C},
number = {2},
pages = {1432--1440},
title = {{Highly Asymmetric Optical Properties of $\beta-\textrm{Ga}_{2}\textrm{O}_{3}$ as Probed by Linear and Nonlinear Optical Excitation Spectroscopy}},
volume = {125},
year = {2021},
}

@article{Dohy1982,
author = {Dohy, D. and Lucazeau, G. and Revcolevschi, A.},
Xdoi = {10.1016/0022-4596(82)90274-2},
issn = {00224596},
journal = {J. Solid State Chem.},
month = {nov},
number = {2},
pages = {180--192},
title = {Raman spectra and valence force field of single-crystalline $\beta-\textrm{Ga}_{2}\textrm{O}_{3}$},
url = {https://linkinghub.elsevier.com/retrieve/pii/0022459682902742},
volume = {45},
year = {1982},
}

@article{Fiedler2020,
author = {Fiedler, A. and Ramsteiner, M. and Galazka, Z. and Irmscher, K.},
Xdoi = {10.1063/5.0024494},
issn = {0003-6951},
journal = {Appl. Phys. Lett.},
month = {oct},
number = {15},
pages = {152107},
publisher = {AIP Publishing LLC},
title = {{Raman scattering in heavily donor doped $\beta-\textrm{Ga}_{2}\textrm{O}_{3}$}},
url = {http://aip.scitation.org/Xdoi/10.1063/5.0024494},
volume = {117},
year = {2020},
}

@article{Furthmueller2016,
author = {Furthm{\"{u}}ller, J. and Bechstedt, F.},
Xdoi = {10.1103/PhysRevB.93.115204},
issn = {2469-9950},
journal = {Phys. Rev. B},
month = {mar},
number = {11},
pages = {115204},
title = {Quasiparticle bands and spectra of $\textrm{Ga}_{2}\textrm{O}_{3}$ polymorphs},
url = {https://link.aps.org/Xdoi/10.1103/PhysRevB.93.115204},
volume = {93},
year = {2016}
}

@article{Galazka2017,
author = {Galazka, Zbigniew and Uecker, Reinhard and Klimm, Detlef and Irmscher, Klaus and Naumann, Martin and Pietsch, Mike and Kwasniewski, Albert and Bertram, Rainer and Ganschow, Steffen and Bickermann, Matthias},
Xdoi = {10.1149/2.0021702JSS/XML},
issn = {2162-8769},
journal = {ECS J. Solid State Sci. Technol.},
month = {sep},
number = {2},
pages = {Q3007--Q3011},
publisher = {The Electrochemical Society},
title = {{ Scaling-Up of Bulk $\beta-\mathrm{Ga}_{2}\mathrm{O}_{3}$ Single Crystals by the Czochralski Method }},
url = {https://iopscience.iop.org/article/10.1149/2.0021702jss https://iopscience.iop.org/article/10.1149/2.0021702jss/meta},
volume = {6},
year = {2017}
}

@article{Galazka2021,
author = {Galazka, Zbigniew and Ganschow, Steffen and Irmscher, Klaus and Klimm, Detlef and Albrecht, Martin and Schewski, Robert and Pietsch, Mike and Schulz, Tobias and Dittmar, Andrea and Kwasniewski, Albert and Grueneberg, Raimund and Anooz, Saud Bin and Popp, Andreas and Juda, Uta and Hanke, Isabelle M. and Schroeder, Thomas and Bickermann, Matthias},
Xdoi = {10.1016/j.pcrysgrow.2020.100511},
issn = {09608974},
journal = {Prog. Cryst. Growth Charact. Mater.},
month = {feb},
number = {1},
pages = {100511},
publisher = {Elsevier Ltd},
title = {{Bulk single crystals of $\beta-\mathrm{Ga}_{2}\mathrm{O}_{3}$ and Ga-based spinels as ultra-wide bandgap transparent semiconducting oxides}},
url = {https://linkinghub.elsevier.com/retrieve/pii/S0960897420300383},
volume = {67},
year = {2021}
}

@article{Galazka2022,
author = {Galazka, Zbigniew},
Xdoi = {10.1063/5.0076962},
issn = {0021-8979},
journal = {J. Appl. Phys.},
month = {jan},
number = {3},
pages = {031103},
publisher = {AIP Publishing LLC},
title = {{Growth of bulk $\beta-\mathrm{Ga}_{2}\mathrm{O}_{3}$ single crystals by the Czochralski method}},
url = {https://aip.scitation.org/Xdoi/10.1063/5.0076962},
volume = {131},
year = {2022}
}

@article{Ghosh2016,
author = {Ghosh, Krishnendu and Singisetti, Uttam},
Xdoi = {10.1063/1.4961308},
isbn = {7166451017},
issn = {00036951},
journal = {Appl. Phys. Lett.},
number = {7},
title = {{Ab initio calculation of electron-phonon coupling in monoclinic $\beta-\mathrm{Ga}_{2}\mathrm{O}_{3}$ crystal}},
url = {http://dx.Xdoi.org/10.1063/1.4961308},
volume = {109},
year = {2016}
}

@article{Gopalan2020,
author = {Gopalan, Prashanth and Knight, Sean and Chanana, Ashish and Stokey, Megan and Ranga, Praneeth and Scarpulla, Michael A. and Krishnamoorthy, Sriram and Darakchieva, Vanya and Galazka, Zbigniew and Irmscher, Klaus and Fiedler, Andreas and Blair, Steve and Schubert, Mathias and Sensale-Rodriguez, Berardi},
Xdoi = {10.1063/5.0031464},
issn = {00036951},
journal = {Appl. Phys. Lett.},
number = {25},
publisher = {AIP Publishing LLC},
title = {{The anisotropic quasi-static permittivity of single-crystal $\beta-\mathrm{Ga}_{2}\mathrm{O}_{3}$ measured by terahertz spectroscopy}},
volume = {117},
year = {2020}
}

@article{Green2022,
author = {Green, Andrew J. and Speck, James and Xing, Grace and Moens, Peter and Allerstam, Fredrik and Gumaelius, Krister and Neyer, Thomas and Arias-Purdue, Andrea and Mehrotra, Vivek and Kuramata, Akito and Sasaki, Kohei and Watanabe, Shinya and Koshi, Kimiyoshi and Blevins, John and Bierwagen, Oliver and Krishnamoorthy, Sriram and Leedy, Kevin and Arehart, Aaron R. and Neal, Adam T. and Mou, Shin and Ringel, Steven A. and Kumar, Avinash and Sharma, Ankit and Ghosh, Krishnendu and Singisetti, Uttam and Li, Wenshen and Chabak, Kelson and Liddy, Kyle and Islam, Ahmad and Rajan, Siddharth and Graham, Samuel and Choi, Sukwon and Cheng, Zhe and Higashiwaki, Masataka},
Xdoi = {10.1063/5.0060327},
issn = {2166-532X},
journal = {APL Mater.},
month = {feb},
number = {2},
pages = {029201},
publisher = {AIP Publishing, LLC},
title = {{$\beta$-Gallium oxide power electronics}},
url = {https://Xdoi.org/10.1063/5.0060327 https://aip.scitation.org/Xdoi/10.1063/5.0060327},
volume = {10},
year = {2022}
}

@article{Grundmann2017,
author = {Grundmann, Marius and Sturm, Chris and Kranert, Christian and Richter, Steffen and Schmidt-Grund, R{\"{u}}diger and Deparis, Christianne and Z{\'{u}}{\~{n}}iga-P{\'{e}}rez, Jes{\'{u}}s},
Xdoi = {10.1002/pssr.201600295},
issn = {18626254},
journal = {Phys. Status Solidi Rapid Res. Lett.},
month = {jan},
number = {1},
pages = {1600295},
title = {{Optically anisotropic media: New approaches to the dielectric function, singular axes, microcavity modes and Raman scattering intensities}},
url = {https://onlinelibrary.wiley.com/Xdoi/10.1002/pssr.201600295},
volume = {11},
year = {2017}
}

@article{Higashiwaki,
author = {Higashiwaki, Masataka and Sasaki, Kohei and Kuramata, Akito and Masui, Takekazu and Yamakoshi, Shigenobu},
Xdoi = {10.1002/pssa.201330197},
issn = {18626300},
journal = {Phys. Status Solidi A},
month = {jan},
number = {1},
pages = {21--26},
title = {Development of gallium oxide power devices},
url = {http://Xdoi.wiley.com/10.1002/pssa.201330197},
volume = {211},
year = {2014}
}

@article{Higashiwaki2,
author = {Higashiwaki, Masataka and Sasaki, Kohei and Kuramata, Akito and Masui, Takekazu and Yamakoshi, Shigenobu},
Xdoi = {10.1063/1.3674287},
issn = {0003-6951},
journal = {Appl. Phys. Lett.},
month = {jan},
number = {1},
pages = {013504},
title = {Gallium oxide ($\mathrm{Ga}_{2}\mathrm{O}_{3}$) metal-semiconductor field-effect transistors on single-crystal $\beta-\mathrm{Ga}_{2}\mathrm{O}_{3}$ (010) substrates},
url = {http://aip.scitation.org/Xdoi/10.1063/1.3674287},
volume = {100},
year = {2012}
}

@article{Higashiwaki2017,
author = {Higashiwaki, Masataka and Kuramata, Akito and Murakami, Hisashi and Kumagai, Yoshinao},
Xdoi = {10.1088/1361-6463/aa7aff},
issn = {0022-3727},
journal = {J. Phys. D. Appl. Phys.},
month = {aug},
number = {33},
pages = {333002},
publisher = {IOP Publishing},
title = {{State-of-the-art technologies of gallium oxide power devices}},
url = {https://iopscience.iop.org/article/10.1088/1361-6463/aa7aff},
volume = {50},
year = {2017}
}

@article{Hildebrandt2021,
author = {Hildebrandt, R. and Sturm, C. and Wieneke, M. and Dadgar, A. and Grundmann, M.},
Xdoi = {10.1063/5.0060198},
issn = {0003-6951},
journal = {Appl. Phys. Lett.},
month = {sep},
number = {12},
pages = {121109},
publisher = {AIP Publishing LLC},
title = {{Raman tensor determination of transparent uniaxial crystals and their thin films—a-plane GaN as exemplary case}},
url = {https://aip.scitation.org/Xdoi/10.1063/5.0060198},
volume = {119},
year = {2021}
}

@article{Janzen2021_beta,
author = {Janzen, Benjamin M. and Mazzolini, Piero and Gillen, Roland and Falkenstein, Andreas and Martin, Manfred and Tornatzky, Hans and Maultzsch, Janina and Bierwagen, Oliver and Wagner, Markus R.},
Xdoi = {10.1039/D0TC04101G},
issn = {2050-7526},
journal = {J. Mater. Chem. C},
month = {jan},
pages = {2311--2320},
title = {Isotopic study of Raman active phonon modes in $\beta-\mathrm{Ga}_{2}\mathrm{O}_{3}$},
url = {https://pubs.rsc.org/en/content/articlehtml/2021/tc/d0tc04101g https://pubs.rsc.org/en/content/articlelanding/2021/tc/d0tc04101g http://xlink.rsc.org/?Xdoi=D0TC04101G},
volume = {9},
year = {2021}
}

@article{Janzen2022,
  title = {First- and second-order Raman spectroscopy of monoclinic $\ensuremath{\beta}\ensuremath{-}{\mathrm{Ga}}_{2}{\mathrm{O}}_{3}$},
  author = {Janzen, Benjamin M. and Gillen, Roland and Galazka, Zbigniew and Maultzsch, Janina and Wagner, Markus R.},
  journal = {Phys. Rev. Mater.},
  volume = {6},
  issue = {5},
  pages = {054601},
  numpages = {11},
  year = {2022},
  month = {May},
  publisher = {American Physical Society},
  Xdoi = {10.1103/PhysRevMaterials.6.054601},
  url = {https://link.aps.org/Xdoi/10.1103/PhysRevMaterials.6.054601}
}

@article{Kokubuna2007,
author = {Kokubun, Yoshihiro and Miura, Kasumi and Endo, Fumie and Nakagomi, Shinji},
Xdoi = {10.1063/1.2432946},
issn = {0003-6951},
journal = {Appl. Phys. Lett.},
month = {jan},
number = {3},
pages = {031912},
title = {Sol-gel prepared $\beta-\mathrm{Ga}_{2}\mathrm{O}_{3}$ thin films for ultraviolet photodetectors},
url = {http://aip.scitation.org/Xdoi/10.1063/1.2432946},
volume = {90},
year = {2007}
}

@article{Kosc2020,
author = {Kosc, T. Z. and Huang, H. and Kessler, T. J. and Negres, R. A. and Demos, S. G.},
Xdoi = {10.1038/s41598-020-73163-4},
isbn = {0123456789},
issn = {2045-2322},
journal = {Sci. Rep.},
month = {dec},
number = {1},
pages = {16283},
pmid = {33004935},
publisher = {Nature Publishing Group UK},
title = {{Determination of the Raman polarizability tensor in the optically anisotropic crystal potassium dihydrogen phosphate and its deuterated analog}},
url = {https://Xdoi.org/10.1038/s41598-020-73163-4 https://www.nature.com/articles/s41598-020-73163-4},
volume = {10},
year = {2020}
}

@Article{Kranert2016SciRep,
  author          = {Kranert, Christian and Sturm, Chris and Schmidt-Grund, R{\"{u}}diger and Grundmann, Marius},
  title           = {{Raman tensor elements of $\beta$-Ga2O3}},
  journal         = {Sci. Rep.},
  year            = {2016},
  volume          = {6},
  number          = {1},
  pages           = {35964},
  month           = {dec},
  issn            = {2045-2322},
  Xdoi             = {10.1038/srep35964},
  url             = {www.nature.com/scientificreports/ http://www.nature.com/articles/srep35964},
}

@Article{Kranert2016PRL,
  author   = {Kranert, Christian and Sturm, Chris and Schmidt-Grund, R{\"{u}}diger and Grundmann, Marius},
  title    = {{Raman Tensor Formalism for Optically Anisotropic Crystals}},
  journal  = {Phys. Rev. Lett.},
  year     = {2016},
  volume   = {116},
  number   = {12},
  pages    = {127401},
  month    = {mar},
  issn     = {0031-9007},
  Xdoi      = {10.1103/PhysRevLett.116.127401},
  url      = {https://journals.aps.org/prl/pdf/10.1103/PhysRevLett.116.127401 https://link.aps.org/Xdoi/10.1103/PhysRevLett.116.127401},
}

@article{Liu2007,
author = {Liu, Bo and Gu, Mu and Liu, Xiaolin},
Xdoi = {10.1063/1.2800792},
issn = {0003-6951},
journal = {Appl. Phys. Lett.},
month = {oct},
number = {17},
pages = {172102},
title = {Lattice dynamical, dielectric, and thermodynamic properties of $\beta-\mathrm{Ga}_{2}\mathrm{O}_{3}$ from first principles},
url = {http://aip.scitation.org/Xdoi/10.1063/1.2800792},
volume = {91},
year = {2007}
}

@article{Look2019,
author = {Look, David C. and Leedy, Kevin D.},
Xdoi = {10.1038/s41598-018-38419-0},
isbn = {4159801838},
issn = {2045-2322},
journal = {Sci. Rep.},
month = {feb},
number = {1},
pages = {1290},
pmid = {30718714},
publisher = {Springer US},
title = {{Classical and quantum conductivity in $\beta-\mathrm{Ga}_{2}\mathrm{O}_{3}$}},
url = {http://dx.Xdoi.org/10.1038/s41598-018-38419-0 http://www.ncbi.nlm.nih.gov/pubmed/30718714 http://www.pubmedcentral.nih.gov/articlerender.fcgi?artid=PMC6362195},
volume = {9},
year = {2019}
}

@article{Machon2006,
author = {Machon, Denis and McMillan, Paul F. and Xu, Bin and Dong, Jianjun},
Xdoi = {10.1103/PhysRevB.73.094125},
issn = {1098-0121},
journal = {Phys. Rev. B},
month = {mar},
number = {9},
pages = {094125},
title = {High-pressure study of the $\beta-\ \mathrm{to}-\alpha$ transition in $\textrm{Ga}_{2}\textrm{O}_{3}$},
url = {https://link.aps.org/Xdoi/10.1103/PhysRevB.73.094125},
volume = {73},
year = {2006}
}

@article{Matsumoto1974,
author = {Matsumoto, Takashi and Aoki, Masaharu and Kinoshita, Akira and Aono, Tomoyoshi},
Xdoi = {10.1143/JJAP.13.1578},
issn = {0021-4922},
journal = {Jpn. J. Appl. Phys.},
month = {oct},
number = {10},
pages = {1578--1582},
publisher = {IOP Publishing},
title = {{Absorption and Reflection of Vapor Grown Single Crystal Platelets of $\beta-\mathrm{Ga}_{2}\mathrm{O}_{3}$}},
url = {https://iopscience.iop.org/article/10.1143/JJAP.13.1578 https://iopscience.iop.org/article/10.1143/JJAP.13.1578/meta},
volume = {13},
year = {1974}
}

@article{Mengle2019,
author = {Mengle, K. A. and Kioupakis, E.},
Xdoi = {10.1063/1.5055238},
issn = {2158-3226},
journal = {AIP Adv.},
month = {jan},
number = {1},
pages = {015313},
publisher = {American Institute of Physics Inc.},
title = {{Vibrational and electron-phonon coupling properties of $\beta-\mathrm{Ga}_{2}\mathrm{O}_{3}$ from first-principles calculations: Impact on the mobility and breakdown field}},
url = {http://aip.scitation.org/Xdoi/10.1063/1.5055238},
volume = {9},
year = {2019}
}

@article{Mock2017,
author = {Mock, Alyssa and Korlacki, Rafa{\l} and Briley, Chad and Darakchieva, Vanya and Monemar, Bo and Kumagai, Yoshinao and Goto, Ken and Higashiwaki, Masataka and Schubert, Mathias},
Xdoi = {10.1103/PhysRevB.96.245205},
issn = {2469-9950},
journal = {Phys. Rev. B},
month = {dec},
number = {24},
pages = {245205},
title = {{Band-to-band transitions, selection rules, effective mass, and excitonic contributions in monoclinic \mbox{$\beta$-\mathrm{Ga}_{2}\mathrm{O}_{3}}}},
url = {https://link.aps.org/Xdoi/10.1103/PhysRevB.96.245205},
volume = {96},
year = {2017}
}

@article{Momma2011,
author = {Momma, Koichi and Izumi, Fujio},
Xdoi = {10.1107/S0021889811038970},
xissn = {00218898},
journal = {J. Appl. Crystallogr.},
month = {dec},
number = {6},
pages = {1272--1276},
publisher = {International Union of Crystallography},
title = {{VESTA 3 for three-dimensional visualization of crystal, volumetric and morphology data}},
volume = {44},
year = {2011},
xnote = {\url{http://scripts.iucr.org/cgi-bin/paper?db5098}},
}

@article{Onuma2014,
author = {Onuma, T. and Fujioka, S. and Yamaguchi, T. and Itoh, Y. and Higashiwaki, M. and Sasaki, K. and Masui, T. and Honda, T.},
Xdoi = {10.1016/j.jcrysgro.2013.12.061},
issn = {00220248},
journal = {J. Cryst. Growth},
month = {sep},
pages = {330--333},
publisher = {Elsevier},
title = {Polarized Raman spectra in $\beta-\mathrm{Ga}_{2}\mathrm{O}_{3}$ single crystals},
url = {http://dx.Xdoi.org/10.1016/j.jcrysgro.2013.12.061 https://linkinghub.elsevier.com/retrieve/pii/S0022024814000475},
volume = {401},
year = {2014}
}

@article{Onuma2015,
author = {Onuma, Takeyoshi and Saito, Shingo and Sasaki, Kohei and Masui, Tatekazu and Yamaguchi, Tomohiro and Honda, Tohru and Higashiwaki, Masataka},
Xdoi = {10.7567/JJAP.54.112601},
issn = {0021-4922},
journal = {Jpn. J. Appl. Phys.},
month = {nov},
number = {11},
pages = {112601},
title = {{Valence band ordering in $\beta-\mathrm{Ga}_{2}\mathrm{O}_{3}$ studied by polarized transmittance and reflectance spectroscopy}},
url = {https://iopscience.iop.org/article/10.7567/JJAP.54.112601},
volume = {54},
year = {2015}
}

@article{Onuma2016,
author = {Onuma, T. and Saito, S. and Sasaki, K. and Goto, K. and Masui, T. and Yamaguchi, T. and Honda, T. and Kuramata, A. and Higashiwaki, M.},
Xdoi = {10.1063/1.4943175},
issn = {0003-6951},
journal = {Appl. Phys. Lett.},
month = {mar},
number = {10},
pages = {101904},
title = {{Temperature-dependent exciton resonance energies and their correlation with IR-active optical phonon modes in $\beta-\mathrm{Ga}_{2}\mathrm{O}_{3}$ single crystals}},
url = {http://dx.Xdoi.org/10.1063/1.4943175 http://aip.scitation.org/Xdoi/10.1063/1.4943175},
volume = {108},
year = {2016}
}

@article{Onuma2016b,
author = {Onuma, Takeyoshi and Saito, Shingo and Sasaki, Kohei and Masui, Tatekazu and Yamaguchi, Tomohiro and Honda, Tohru and Kuramata, Akito and Higashiwaki, Masataka},
Xdoi = {10.7567/JJAP.55.1202B2},
issn = {13474065},
journal = {Jpn. J. Appl. Phys.},
number = {12},
title = {{Spectroscopic ellipsometry studies on $\beta-\mathrm{Ga}_{2}\mathrm{O}_{3}$ films and single crystal}},
volume = {55},
year = {2016}
}

@article{Oshima2008,
author = {Oshima, Takayoshi and Okuno, Takeya and Arai, Naoki and Suzuki, Norihito and Ohira, Shigeo and Fujita, Shizuo},
Xdoi = {10.1143/APEX.1.011202},
issn = {1882-0778},
journal = {Appl. Phys. Express},
month = {jan},
number = {1},
pages = {011202},
title = {Vertical Solar-Blind Deep-Ultraviolet Schottky Photodetectors Based on $\beta-\mathrm{Ga}_{2}\mathrm{O}_{3}$ Substrates},
url = {https://scinapse.io/papers/2006465672 https://iopscience.iop.org/article/10.1143/APEX.1.011202},
volume = {1},
year = {2008}
}

@article{Parisini2018,
author = {Parisini, A. and Ghosh, K. and Singisetti, U. and Fornari, R.},
Xdoi = {10.1088/1361-6641/aad5cd},
issn = {0268-1242},
journal = {Semicond. Sci. Technol.},
month = {oct},
number = {10},
pages = {105008},
publisher = {IOP Publishing},
title = {{Assessment of phonon scattering-related mobility in $\beta-\mathrm{Ga}_{2}\mathrm{O}_{3}$}},
url = {https://iopscience.iop.org/article/10.1088/1361-6641/aad5cd},
volume = {33},
year = {2018}
}

@article{Pearton,
author = {Pearton, S. J. and Yang, Jiancheng and Cary, Patrick H. and Ren, F. and Kim, Jihyun and Tadjer, Marko J. and Mastro, Michael A.},
Xdoi = {10.1063/1.5006941},
issn = {1931-9401},
journal = {Appl. Phys. Rev.},
month = {mar},
number = {1},
pages = {011301},
title = {A review of $\mathrm{Ga}_{2}\mathrm{O}_{3}$ materials, processing, and devices},
url = {http://dx.Xdoi.org/10.1063/1.5006941 http://aip.scitation.org/Xdoi/10.1063/1.5006941},
volume = {5},
year = {2018}
}

@article{Pearton2018,
author = {Pearton, S. J. and Ren, Fan and Tadjer, Marko and Kim, Jihyun},
Xdoi = {10.1063/1.5062841},
issn = {0021-8979},
journal = {J. Appl. Phys.},
month = {dec},
number = {22},
pages = {220901},
title = {{Perspective: $\mathrm{Ga}_{2}\mathrm{O}_{3}$ for ultra-high power rectifiers and \textrm{MOSFETS}}},
url = {http://dx.Xdoi.org/10.1063/1.5062841 http://aip.scitation.org/Xdoi/10.1063/1.5062841},
volume = {124},
year = {2018}
}

@article{Ratnaparkhe2017,
author = {Ratnaparkhe, Amol and Lambrecht, Walter R. L.},
Xdoi = {10.1063/1.4978668},
issn = {0003-6951},
journal = {Appl. Phys. Lett.},
month = {mar},
number = {13},
pages = {132103},
title = {{Quasiparticle self-consistent GW band structure of $\mathrm{Ga}_{2}\mathrm{O}_{3}$ and the anisotropy of the absorption onset}},
url = {http://aip.scitation.org/Xdoi/10.1063/1.4978668},
volume = {110},
year = {2017}
}

@article{Ricci2016,
author = {Ricci, F. and Boschi, F. and Baraldi, A. and Filippetti, A. and Higashiwaki, M. and Kuramata, A. and Fiorentini, V. and Fornari, R.},
Xdoi = {10.1088/0953-8984/28/22/224005},
issn = {1361648X},
journal = {J. Phys. Condens. Matter},
number = {22},
pages = {224005},
publisher = {IOP Publishing},
title = {{Theoretical and experimental investigation of optical absorption anisotropy in $\mathrm{Ga}_{2}\mathrm{O}_{3}$}},
url = {http://dx.Xdoi.org/10.1088/0953-8984/28/22/224005},
volume = {28},
year = {2016}
}

@article{Ritter2019,
author = {Ritter, Jacob R. and Lynn, Kelvin G. and McCluskey, Matthew D.},
Xdoi = {10.1063/1.5129781},
issn = {0021-8979},
journal = {J. Appl. Phys.},
month = {dec},
number = {22},
pages = {225705},
publisher = {AIP Publishing LLC},
title = {{Iridium-related complexes in Czochralski-grown $\beta-\mathrm{Ga}_{2}\mathrm{O}_{3}$}},
url = {http://aip.scitation.org/Xdoi/10.1063/1.5129781},
volume = {126},
year = {2019}
}

@article{Sander2012,
author = {Sander, T. and Eisermann, S. and Meyer, B. K. and Klar, P. J.},
Xdoi = {10.1103/PhysRevB.85.165208},
issn = {1098-0121},
journal = {Phys. Rev. B},
month = {apr},
number = {16},
pages = {165208},
title = {{Raman tensor elements of wurtzite ZnO}},
url = {https://link.aps.org/Xdoi/10.1103/PhysRevB.85.165208},
volume = {85},
year = {2012}
}

@article{Schubert2016,
XarchivePrefix = {arXiv},
XarxivId = {1512.08590},
author = {Schubert, M. and Korlacki, R. and Knight, S. and Hofmann, T. and Sch{\"{o}}che, S. and Darakchieva, V. and Janz{\'{e}}n, E. and Monemar, B. and Gogova, D. and Thieu, Q.-T. and Togashi, R. and Murakami, H. and Kumagai, Y. and Goto, K. and Kuramata, A. and Yamakoshi, S. and Higashiwaki, M.},
Xdoi = {10.1103/PhysRevB.93.125209},
Xeprint = {1512.08590},
issn = {2469-9950},
journal = {Phys. Rev. B},
month = {mar},
number = {12},
pages = {125209},
title = {Anisotropy, phonon modes, and free charge carrier parameters in monoclinic $\beta-$gallium oxide single crystals},
url = {https://link.aps.org/Xdoi/10.1103/PhysRevB.93.125209},
volume = {93},
year = {2016}
}

@article{Seyidov2022,
author = {Seyidov, Palvan and Ramsteiner, Manfred and Galazka, Zbigniew and Irmscher, Klaus},
Xdoi = {10.1063/1.5081825},
issn = {0021-8979},
journal = {J. Appl. Phys.},
month = {jan},
number = {3},
pages = {035707},
title = {{Resonant electronic Raman scattering from $\mathrm{Ir}^{4+}$ ions in $\beta-\mathrm{Ga}_{2}\mathrm{O}_{3}$}},
url = {https://aip.scitation.org/Xdoi/10.1063/5.0080248},
volume = {131},
year = {2022}
}

@article{Spencer2022,
author = {Spencer, Joseph A. and Mock, Alyssa L. and Jacobs, Alan G. and Schubert, Mathias and Zhang, Yuhao and Tadjer, Marko J.},
Xdoi = {10.1063/5.0078037},
issn = {19319401},
journal = {Appl. Phys. Rev.},
number = {1},
publisher = {AIP Publishing LLC},
title = {{A review of band structure and material properties of transparent conducting and semiconducting oxides: $\mathrm{Ga}_{2}\mathrm{O}_{3}$, $\mathrm{Al}_{2}\mathrm{O}_{3}$, $\mathrm{In}_{2}\mathrm{O}_{3}$, $\mathrm{ZnO}$, $\mathrm{Sn}\mathrm{O}_{2}$, $\mathrm{CdO}$, $\mathrm{NiO}$, $\mathrm{CuO}$, and $\mathrm{Sc}_{2}\mathrm{O}_{3}$}},
volume = {9},
year = {2022}
}

@Article{Sturm2015,
  author        = {Sturm, C. and Furthm{\"{u}}ller, J. and Bechstedt, F. and Schmidt-Grund, R. and Grundmann, M.},
  title         = {{Dielectric tensor of monoclinic $\mathrm{Ga}_{2}\mathrm{O}_{3}$ single crystals in the spectral range 0.5-8.5 eV}},
  journal       = {APL Materials},
  year          = {2015},
  volume        = {3},
  number        = {10},
  pages         = {0--9},
  issn          = {2166532X},
  XarchivePrefix = {arXiv},
  XarxivId       = {1507.05401},
  Xdoi           = {10.1063/1.4934705},
  Xeprint        = {1507.05401},
  isbn          = {2121042121},
}

@article{Sturm2016,
author = {Sturm, C. and Schmidt-Grund, R. and Kranert, C. and Furthm{\"{u}}ller, J. and Bechstedt, F. and Grundmann, M.},
Xdoi = {10.1103/PhysRevB.94.035148},
issn = {2469-9950},
journal = {Phys. Rev. B},
month = {jul},
number = {3},
pages = {035148},
publisher = {American Physical Society},
title = {{Dipole analysis of the dielectric function of color dispersive materials: Application to monoclinic \mbox{$\mathrm{Ga}_{2}\mathrm{O}_{3}$}} },
url = {https://link.aps.org/Xdoi/10.1103/PhysRevB.94.035148},
volume = {94},
year = {2016}
}

@article{Sturm2016b,
XarchivePrefix = {arXiv},
XarxivId = {1601.03760},
author = {Sturm, Chris and Grundmann, Marius},
Xdoi = {10.1103/PhysRevA.93.053839},
Xeprint = {1601.03760},
issn = {2469-9926},
journal = {Phys. Rev. A},
month = {may},
number = {5},
pages = {053839},
title = {{Singular optical axes in biaxial crystals and analysis of their spectral dispersion effects in \textbf{\mbox{$\mathrm{Ga}_{2}\mathrm{O}_{3}$}} }},
url = {https://link.aps.org/Xdoi/10.1103/PhysRevA.93.053839},
volume = {93},
year = {2016}
}

@article{Ueda1997,
author = {Ueda, Naoyuki and Hosono, Hideo and Waseda, Ryuta and Kawazoe, Hiroshi},
Xdoi = {10.1063/1.119693},
issn = {0003-6951},
journal = {Appl. Phys. Lett.},
month = {aug},
number = {7},
pages = {933--935},
title = {Anisotropy of electrical and optical properties in $\beta-\mathrm{Ga}_{2}\mathrm{O}_{3}$ single crystals},
url = {http://aip.scitation.org/Xdoi/10.1063/1.119693},
volume = {71},
year = {1997}
}

@article{Varley2015,
author = {Varley, Joel B. and Schleife, Andr{\'{e}}},
Xdoi = {10.1088/0268-1242/30/2/024010},
issn = {0268-1242},
journal = {Semicond. Sci. Technol.},
month = {feb},
number = {2},
pages = {024010},
publisher = {IOP Publishing},
title = {{Bethe–Salpeter calculation of optical-absorption spectra of $\mathrm{In}_{2}\mathrm{O}_{3}$ and $\mathrm{Ga}_{2}\mathrm{O}_{3}$}},
url = {https://iopscience.iop.org/article/10.1088/0268-1242/30/2/024010},
volume = {30},
year = {2015}
}

@article{Villora2002,
author = {Villora, E.G. and Morioka, Y. and Atou, T. and Sugawara, T. and Kikuchi, M. and Fukuda, T.},
Xdoi = {10.1002/1521-396X(200209)193:1<187::AID-PSSA187>3.0.CO;2-1},
issn = {0031-8965},
journal = {Phys. Status Solidi A},
month = {sep},
number = {1},
pages = {187--195},
title = {Infrared Reflectance and Electrical Conductivity of $\beta-\mathrm{Ga}_{2}\mathrm{O}_{3}$},
url = {https://onlinelibrary.wiley.com/Xdoi/abs/10.1002/1521-396X(200209)193:1%3C187::AID-PSSA187%3E3.0.CO;2-1},
volume = {193},
year = {2002}
}

@article{VonWenckstern2017,
author = {von Wenckstern, Holger},
Xdoi = {10.1002/aelm.201600350},
issn = {2199160X},
journal = {Adv. Electron. Mater.},
month = {sep},
number = {9},
pages = {1600350},
title = {{$\mathrm{Group-}III$ Sesquioxides: Growth, Physical Properties and Devices}},
url = {https://onlinelibrary.wiley.com/Xdoi/10.1002/aelm.201600350},
volume = {3},
year = {2017}
}

@article{Yao2019,
author = {Yao, Yongzhao and Ishikawa, Yukari and Sugawara, Yoshihiro},
Xdoi = {10.1063/1.5129226},
issn = {10897550},
journal = {J. Appl. Phys.},
number = {20},
pages = {205106},
publisher = {AIP Publishing LLC},
title = {{X-ray diffraction and Raman characterization of $\beta-\mathrm{Ga}_{2}\mathrm{O}_{3}$ single crystal grown by edge-defined film-fed growth method}},
volume = {126},
year = {2019}
}

@article{Yamaguchi2004,
author = {Yamaguchi, Kenji},
Xdoi = {10.1016/j.ssc.2004.07.030},
issn = {00381098},
journal = {Solid State Commun.},
month = {sep},
number = {12},
pages = {739--744},
publisher = {Elsevier Ltd},
title = {{First principles study on electronic structure of $\beta$-Ga 2O3}},
volume = {131},
year = {2004}
}

@article{Zhang2021optical,
author = {Zhang, Naiji and Kislyakov, Ivan M. and Xia, Changtai and Qi, Hongji and Wang, Jun and Mohamed, H. F.},
Xdoi = {10.1364/OE.427021},
issn = {1094-4087},
journal = {Opt. Express},
month = {jun},
number = {12},
pages = {18587},
pmid = {34154112},
title = {{Anisotropic luminescence and third-order electric susceptibility of $\mathrm{Mg}$-doped gallium oxide under the half-bandgap edge}},
url = {https://opg.optica.org/abstract.cfm?URI=oe-29-12-18587},
volume = {29},
year = {2021}
}

@article{Zhang2021_beta,
author = {Zhang, Kun and Xu, Zongwei and Zhang, Shengnan and Wang, Hong and Cheng, Hongjuan and Hao, Jianmin and Wu, Jintong and Fang, Fengzhou},
Xdoi = {10.1016/j.physb.2020.412624},
issn = {09214526},
journal = {Physica B Condens. Matter},
month = {jan},
pages = {412624},
publisher = {North-Holland},
title = {{Raman and photoluminescence properties of un-/ion-doped $\beta-\mathrm{Ga}_{2}\mathrm{O}_{3}$ single-crystals prepared by edge-defined film-fed growth method}},
url = {https://linkinghub.elsevier.com/retrieve/pii/S0921452620306165},
volume = {600},
year = {2021}
}

@article{Zhang2021_T-dpendent,
author = {Zhang, Kun and Xu, Zongwei and Zhao, Junlei and Wang, Hong and Hao, Jianmin and Zhang, Shengnan and Cheng, Hongjuan and Dong, Bing},
Xdoi = {10.1016/j.jallcom.2021.160665},
issn = {09258388},
journal = {J. Alloys Compd.},
pages = {160665},
publisher = {Elsevier},
title = {{Temperature-dependent Raman and photoluminescence of $\beta-\mathrm{Ga}_{2}\mathrm{O}_{3}$ doped with shallow donors and deep acceptors impurities}},
url = {https://Xdoi.org/10.1016/j.jallcom.2021.160665},
volume = {881},
year = {2021}
}

@article{Zhang2022,
author = {Zhang, Kun and Xu, Zongwei and Zhao, Junlei and Wang, Hong and Hao, Jianmin and Zhang, Shengnan and Cheng, Hongjuan and Dong, Bing},
Xdoi = {10.1016/j.apsusc.2022.152426},
issn = {01694332},
journal = {Appl. Surf. Sci.},
month = {apr},
pages = {152426},
publisher = {Elsevier B.V.},
title = {{Anisotropies of angle-resolved polarized Raman response identifying in low miller index $\beta-\mathrm{Ga}_{2}\mathrm{O}_{3}$ single crystal}},
url = {https://linkinghub.elsevier.com/retrieve/pii/S0169433222000095},
volume = {581},
year = {2022}
}

@article{Grundmann2016,
author = {Grundmann, Marius and Sturm, Chris and Kranert, Christian and Richter, Steffen and Schmidt-Grund, R{\"{u}}diger and Deparis, Christianne and Z{\'{u}}{\~{n}}iga-P{\'{e}}rez, Jes{\'{u}}s},
Xdoi = {10.1002/pssr.201600295},
issn = {18626254},
journal = {Phys. status solidi - Rapid Res. Lett.},
month = {jan},
number = {1},
pages = {1600295},
title = {{Optically anisotropic media: New approaches to the dielectric function, singular axes, microcavity modes and Raman scattering intensities}},
url = {https://onlinelibrary.wiley.com/Xdoi/10.1002/pssr.201600295},
volume = {11},
year = {2016}
}

% \end{document}

\clearpage
\newpage
\setcounter{page}{1}
\setcounter{figure}{0}
\setcounter{section}{0}
\setcounter{table}{0}
\setcounter{equation}{0}
\renewcommand{\thepage}{S\arabic{page}}
\renewcommand{\thetable}{S\arabic{table}}
\renewcommand{\thefigure}{S\arabic{figure}}
\renewcommand{\theequation}{S\arabic{equation}}
\renewcommand{\thesection}{S\arabic{section}}

\onecolumngrid
\section*{}
\begin{center}
    \large Supplemental material to the article:\\[1ex]
 Complete Raman Tensor Determination in Birefringent \texorpdfstring{\betagao{}}{β-Ga₂O₃} by Single-Stage \\
 Hyperspectral Analysis of Polarization Angle-Resolved Raman Spectra
\end{center}

\newcommand{\Ra}{\begin{pmatrix}0 & 0 & 1 \\ 0.961 & -0.278 & 0 \\ 0.278 & 0.961 & 0\end{pmatrix}}
\newcommand{\Rb}{\begin{pmatrix}1 & 0 & 0 \\ 0 & 1 & 0 \\ 0 & 0 & 1\end{pmatrix}}
\newcommand{\Rc}{\begin{pmatrix}0 & 0 & 1 \\ 0.041 & 0.999 & 0 \\ -0.999 & 0.041 & 0\end{pmatrix}}
\newcommand{\Rzwo}{\begin{pmatrix}0 & 0 & 1 \\ -0.800 & -0.600 & 0 \\ 0.600 & -0.800 & 0\end{pmatrix}}

\newcommand{\epsa}{\begin{pmatrix}3.809 & 0 & 0 \\ 0 & 3.765 & 0.031 \\ 0 & 0.031 & 3.678\end{pmatrix}}
\newcommand{\epsb}{\begin{pmatrix}3.727 & 0 & 0 \\ 0 & 3.624 & 0 \\ 0 & 0 & 3.768\end{pmatrix}}
\newcommand{\epsc}{\begin{pmatrix}3.768 & 0 & 0 \\ 0 & 3.624 & -0.004 \\ 0 & -0.004 & 3.727\end{pmatrix}}
\newcommand{\epszwo}{\begin{pmatrix}3.768 & 0 & 0 \\ 0 & 3.690 & -0.049 \\ 0 & -0.049 & 3.661\end{pmatrix}}

\subsection{Conventional coordinate systems in \texorpdfstring{\betagao{}}{β-Ga₂O₃}}

\noindent To understand the theoretical modeling of the selection rules, several coordinate systems are introduced:

\begin{enumerate}
\item The \textit{crystallographic coordinate system} ($a$, $b$, $c$) is defined by $[100] \equiv a$, $[010]\equiv b$, and $[001] \equiv c$ with a monoclinic angle of $\beta=\angle(a,c)=103.8^\circ$ between the $a$- and $c$-axes. The lattice parameters are $a_0=12.21\, \mathrm{\AA}$, $b_0 = 3.04\,\mathrm{\AA}$ and $c=5.80\,\mathrm{\AA}$ \cite{Galazka2021}. 
\item Several equivalent orthonormal \textit{quasi-crystal coordinate systems} ($x$, $y$, $z$) are commonly found in the literature. For one definition $x$, $y$ and $z$ are chosen such that $x\parallel[100]$ and $y\parallel[010]$, while $z$ is defined perpendicular to both axes. Consequently, the angle between $z$ and the crystallographic $c$-axis is $\angle(z,c)=13.8^\circ$. All subsequent coordinate systems are expressed with respect to this basis, which is given by
\begin{equation}
    x=
    \begin{pmatrix}
        1 \\ 0 \\ 0
    \end{pmatrix}, \qquad
        y=
    \begin{pmatrix}
        0 \\ 1 \\ 0
    \end{pmatrix}, \qquad
        z=
    \begin{pmatrix}
        0 \\ 0 \\ 1
    \end{pmatrix}.
\end{equation}
Similarly, permutations of the above definition are equivalently valid, and can be found as the common standard description, depending on the field. Hence, it is worth to note that anisotropic properties may not be readily comparable, such as thermal and elastic properties in \betagao{}~\cite{Kai-thermal}.

\item In the orthonormal \textit{principal axis coordinate system} ($\tilde{x}$, $\tilde{y}$, $\tilde{z}$) are defined by the eigenvectors of the dielectric tensor. For a wavelength of $\lambda = 633$\,nm it is obtained by a rotation of $\alpha=2.33^\circ$ in the $xy$-plane followed by an anticyclic permutation of $x$, $y$ and $z$. Expressed in the above defined quasi-crystal basis, the unit vectors read
\begin{equation}
    \tilde{x}=
    \begin{pmatrix}
        \sin(\alpha) \\ 0 \\ \cos(\alpha)
    \end{pmatrix}, \qquad
        \tilde{y}= 
    \begin{pmatrix}
        \cos(\alpha) \\ 0 \\ -\sin(\alpha)
    \end{pmatrix}, \qquad
        \tilde{z}= 
    \begin{pmatrix}
        0 \\ 1 \\ 0
    \end{pmatrix}.
\end{equation}

\item Finally, four \textit{sample coordinate systems} ($x'_p$, $y'_p$, $z'_p$) with $p\in\{\mathrm{a, b, c, \overline{2}01}\}$ are introduced. Each system is defined such that $z'_p$ is perpendicular to their respective plane, e.\,g.\ $\tilde{z}_{\mathrm{a}} \perp(100)$, which coincides with the direction of light propagation. The axis $x'_p$ is chosen parallel to a crystallographic direction: for the b-plane, $\tilde{x}_{\mathrm{b}}\parallel [100]$, whereas for the other planes $x'_p \parallel [010]$. The remaining axis is then obtained by the cross product $y'_p =z'_p \times x'_p$. The resulting bases, expressed in the quasi-crystal coordinate system, are:

% \begin{align}
%     x'_{\mathrm{a}} &=
%     \begin{pmatrix}
%         0 \\ 1 \\ 0
%     \end{pmatrix}, &
%         y'_{\mathrm{a}} &=
%     \begin{pmatrix} 
%         \cos(\beta) \\ 0 \\ \sin(\beta)
%     \end{pmatrix}, &
%         z'_{\mathrm{a}}&=
%     \begin{pmatrix}
%         \sin(\beta) \\ 0 \\ -\cos(\beta)
%     \end{pmatrix}\\
% %
%     x'_{\mathrm{b}}&=
%     \begin{pmatrix}
%         1 \\ 0 \\ 0
%     \end{pmatrix}, &
%         y'_{\mathrm{b}}&=
%     \begin{pmatrix}
%         0 \\ 0 \\ -1
%     \end{pmatrix}, &
%         z'_{\mathrm{b}}&=
%     \begin{pmatrix}
%         0 \\ 1 \\ 0
%     \end{pmatrix}\\
% %
%     x'_{\mathrm{c}}&=
%     \begin{pmatrix}
%         0 \\ 1 \\ 0
%     \end{pmatrix}, &
%         y'_{\mathrm{c}}&=
%     \begin{pmatrix}
%         1 \\ 0 \\ 0
%     \end{pmatrix}, &
%         z'_{\mathrm{c}}&=
%     \begin{pmatrix}
%         0 \\ 0 \\ -1
%     \end{pmatrix}\\
% %
%     x'_{\mathrm{\overline{2}01}}&=
%     \begin{pmatrix}
%         0 \\ 1 \\ 0
%     \end{pmatrix}, &
%         y'_{\mathrm{\overline{2}01}}&=
%     \begin{pmatrix}
%         -\cos(\gamma) \\ 0 \\ -\sin(\gamma)
%     \end{pmatrix}, &
%         z'_{\mathrm{\overline{2}01}} &=
%     \begin{pmatrix}
%         -\sin(\gamma) \\ 0 \\ \cos(\gamma)
%     \end{pmatrix}, \qquad
%     \gamma = 50.79^\circ.
% \end{align}

\begin{equation}
\begin{aligned}
    \tilde{x}_{\mathrm{a}} = &
    \begin{pmatrix}
        0 \\ 1 \\ 0
    \end{pmatrix}, &
        \tilde{y}_{\mathrm{a}} =&
    \begin{pmatrix} 
        \cos(\beta) \\ 0 \\ \sin(\beta)
    \end{pmatrix}, &
        \tilde{z}_{\mathrm{a}}=&
    \begin{pmatrix}
        \sin(\beta) \\ 0 \\ -\cos(\beta)
    \end{pmatrix}\\
    \tilde{x}_{\mathrm{b}}=&
    \begin{pmatrix}
        \sin(\alpha) \\ 0 \\ \cos(\alpha)
    \end{pmatrix}, &
        \tilde{y}_{\mathrm{b}}=&
    \begin{pmatrix}
        \cos(\alpha) \\ 0 \\ -\sin(\alpha)
    \end{pmatrix}, &
        \tilde{z}_{\mathrm{b}}=&
    \begin{pmatrix}
        0 \\ 1 \\ 0
    \end{pmatrix}\\
    \tilde{x}_{\mathrm{c}}=&
    \begin{pmatrix}
        0 \\ 1 \\ 0
    \end{pmatrix}, &
        \tilde{y}_{\mathrm{c}}=&
    \begin{pmatrix}
        1 \\ 0 \\ 0
    \end{pmatrix}, &
        \tilde{z}_{\mathrm{c}}=&
    \begin{pmatrix}
        0 \\ 0 \\ -1
    \end{pmatrix}\\
    \tilde{x}_{\mathrm{\overline{2}01}}=&
    \begin{pmatrix}
        0 \\ 1 \\ 0
    \end{pmatrix}, \qquad &
        \tilde{y}_{\mathrm{\overline{2}01}}=&
    \begin{pmatrix}
        -\cos(\gamma) \\ 0 \\ -\sin(\gamma)
    \end{pmatrix}, \qquad &
        \tilde{z}_{\mathrm{\overline{2}01}} =&
    \begin{pmatrix}
        -\sin(\gamma) \\ 0 \\ \cos(\gamma)
    \end{pmatrix}, \qquad
    \gamma = 50.79^\circ.
\end{aligned}
\end{equation}
\end{enumerate}

\clearpage
\subsection{Base transformations}
The Raman tensors defined in equation~\ref{eq:Ramantensors} and the dielectric tensor in equation~\ref{eq:dielectrict} are both expressed in the eigenbasis of the dielectric tensor. To obtain their representations in the corresponding measurement geometry, they are transformed using the orthogonal matrix $Q$, whose elements are given by 
\begin{equation}
    Q_{ij} = v'_i \cdot \tilde{v}_j,
\label{eqn:1}
\end{equation}
where $\tilde{v}_j$ denote the basis vectors of the dielectric tensor ($\tilde{x}$, $\tilde{y}$ and $\tilde{z}$), and $v'_i$ denote the basis vectors of the sample system ($x'_p$, $y'_p$ and $z'_p$). The Raman and dielectric tensors in the measurement geometry are then obtained by the basis transformation

\begin{equation}
    R= QRQ^\intercal \qquad \varepsilon = Q\varepsilon Q^\intercal,
\end{equation}
where $Q$ and $\varepsilon$ are shown exemplarily in table~\ref{tab:transform}.
\begin{table*}[hbt]
\caption{Compilation of transformation matrices $Q$ and dielectric tensors $\varepsilon$ representations in the surface systems of the investigated crystal planes.}
\begin{tabular}{ccccc}
\toprule \toprule
& a-plane & b-plane & c-plane & ($\overline{2}$01)-plane \\[1ex]
$Q$: & $\Ra$ & $\Rb$ & $\Rc$ & $\Rzwo$ \\[4ex]
$\varepsilon$: & $\epsa$ & $\epsb$ & $\epsc$ & $\epszwo$ \\
\bottomrule \bottomrule
\label{tab:transform}
\end{tabular}
\end{table*}
 Using these representations of the dielectric tensor and calculating the correction factors $\rho$, $T$ and $Z$ introduced in equation~\ref{eq:efframan}, the explicit selection rules for $\mathrm{A_g}$ and $\mathrm{B_g}$ modes can be obtained for each measurement geometry and are summarized in table~\ref{tab:selectionrules}.
\begin{table*}[hbt]
\caption{Polarization angle dependent selection rules $A(\varphi, r)$, i.\,e.\ {PARRS profiles} for the investigated scattering geometries. Influences of the dielectric tensor described in the extended model for Raman scattering in birefringent media are included in the numerical prefactors.}
\begin{tabular}{c<{\hspace{1em}}c<{\hspace{1em}}c<{\hspace{1em}}}
\toprule \toprule
 & $\mathrm{A_g}$ & $\mathrm{B_g}$ \\
geometry & $ \begin{pmatrix} a & d & 0\\ d & b & 0\\ 0 & 0 & c \end{pmatrix} $ & $ \begin{pmatrix} 0 & 0 & e\\ 0 & 0 & f\\ e & f & 0 \end{pmatrix} $\\
\cmidrule(lr){2-2} \cmidrule(lr){3-3}
\multirow{2}{*}{a-plane} & \rule{0pt}{3ex} $0.970 c^{2} \cos^{4}(\varphi) + \left(0.927 a + 0.073 b - 0.521 d\right)^{2} \sin^{4}(\varphi)$ & $\left(0.952 e - 0.268 f\right)^{2} \sin^{2}(2 \varphi)$ \\\vspace{5pt}
 & \rule{0pt}{3ex} $\frac{1}{4}\left(0.970 c^{2} + \left(0.927 a + 0.073 b - 0.521 d\right)^{2}\right) \sin^{2}(2 \varphi)$ & $\left(0.952 e - 0.268 f\right)^{2} \cos^{2}(2 \varphi)$ \\
 
\multirow{2}{*}{b-plane} & \rule{0pt}{3ex} $0.937 a^{2} \cos^{4}(\varphi) + b^{2} \sin^4(\varphi) + 0.954 d^{2} \sin^2(2 \varphi)$ & $0$ \\\vspace{5pt}
 & \rule{0pt}{3ex} $\left(0.234 a^{2} + 0.25 b^{2} \right) \sin^{2}(2 \varphi) + 0.954 d^{2} \cos^{2}(2 \varphi)$ & $0$ \\
 
\multirow{2}{*}{c-plane} & \rule{0pt}{3ex} $0.913 c^{2} \cos^{4}(\varphi) + \left(0.002 a + 0.998 b + 0.084 d\right)^{2} \sin^{4}(\varphi)$ & $\left(0.040 e + 0.967 f\right)^{2} \sin^{2}(2 \varphi)$ \\\vspace{5pt}
 & \rule{0pt}{3ex} $\frac{1}{4}\left(0.913 c^{2} + \left(0.002 a + 0.998 b + 0.084 d\right)^{2}\right) \sin^{2}(2 \varphi)$ & $\left(0.040 e + 0.967 f\right)^{2} \cos^{2}(2 \varphi)$ \\
 
\multirow{2}{*}{($\overline{2}01$)-plane} & \rule{0pt}{3ex} $0.952 c^{2} \cos^{4}(\varphi) + \left(0.653 a + 0.347 b + 0.952 d\right)^{2} \sin^{4}(\varphi)$ & $\left(0.794 e + 0.579 f\right)^{2} \sin^{2}(2 \varphi)$ \\\vspace{5pt}
 & \rule{0pt}{3ex} $\frac{1}{4}\left(0.952 c^{2} + \left(0.653 a + 0.347 b + 0.952 d\right)^{2}\right) \sin^{2}(2 \varphi)$ & $\left(0.794 e + 0.579 f\right)^{2} \cos^{2}(2 \varphi)$ \\
\bottomrule \bottomrule
\label{tab:selectionrules}
\end{tabular}
\end{table*}

\clearpage
\subsection{Contour plots of the 
\texorpdfstring{$\boldsymbol{(010)}$}{(010)}, 
\texorpdfstring{$\boldsymbol{(001)}$}{(001)} and 
\texorpdfstring{$\boldsymbol{(\overline{2}01)}$}{(201)} planes} 

\begin{figure}[tbh]
\includegraphics[width=0.7\linewidth]{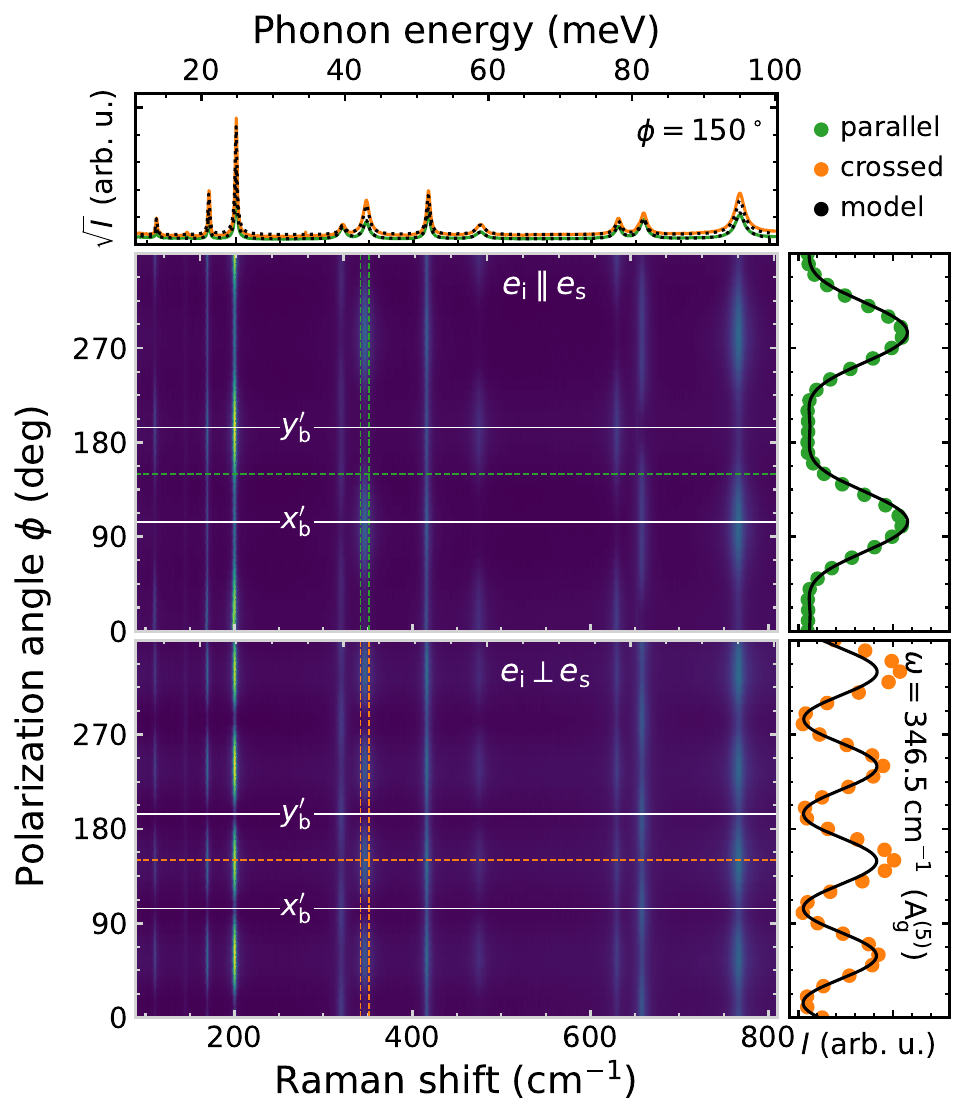}
\caption{Contour plot of the polarization dependent Raman spectra on the b-plane for parallel (middle) and crossed scattering geometries (bottom). Angles, where $y'_{\mathrm{b}} \parallel [100]$ are labeled. In the top panel Raman spectra with a polarization angle of {$\phi=150^\circ$} are depicted, while the panels on the right show the integrated intensity in the range of the $\mathrm{A^{(5)}_g}$ mode and the corresponding slice of the fit. The Raman spectra are depicted with the square root of the intensity for better visualization of small peaks.}
\label{fig:PARRS_b}
\end{figure}

\begin{figure}[tbh]
\includegraphics[width=0.7\linewidth]{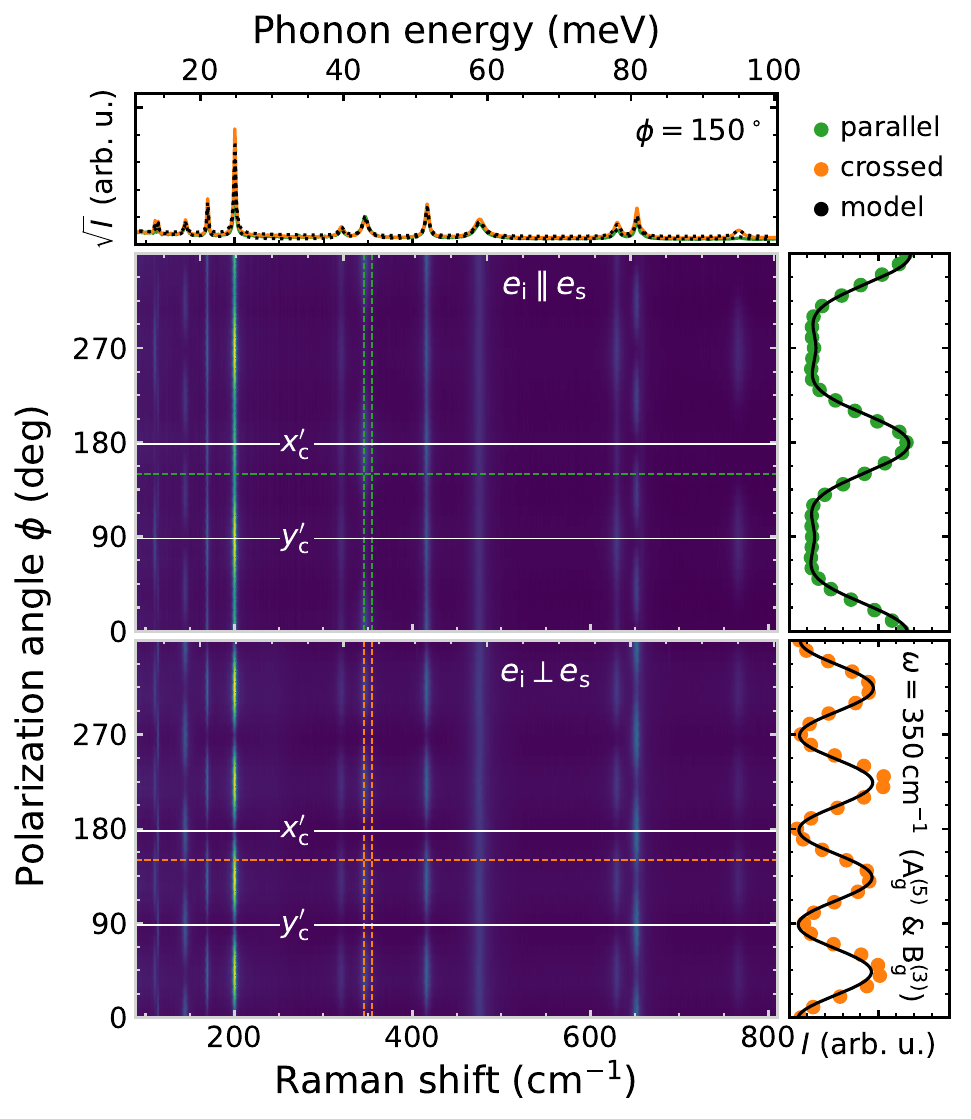}
\caption{Contour plot of the polarization dependent Raman spectra on the c-plane for parallel (middle) and crossed scattering geometries (bottom). Angles, where $x'_{\mathrm{c}} \parallel [010]$ and $y'_{\mathrm{c}} \parallel [100]$ are labeled. In the top panel Raman spectra with a polarization angle of {$\phi=150^\circ$} are depicted, while the panels on the right show the integrated intensity in the range of the $\mathrm{A^{(5)}_g}$/$\mathrm{B^{(3)}_g}$ pair and the corresponding slice of the fit. The Raman spectra are depicted with the square root of the intensity for better visualization of small peaks.}
\label{fig:PARRS_c}
\end{figure}

\begin{figure}[tbh]
\includegraphics[width=0.7\linewidth]{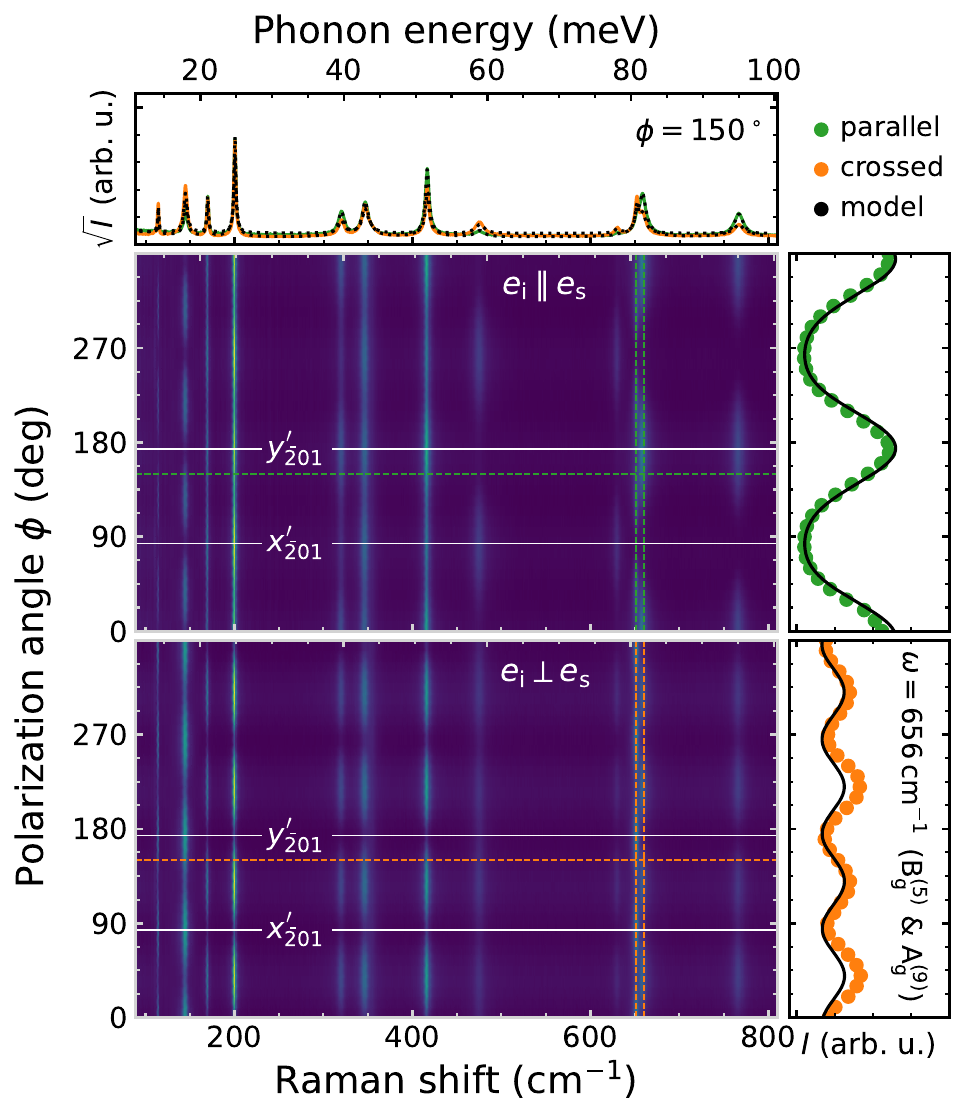}
\caption{Contour plot of the polarization dependent Raman spectra on the $\overline{2}01$-plane for parallel (middle) and crossed scattering geometries (bottom). Angles, where $x'_{\mathrm{\overline{2}01}} \parallel [010]$ are labeled. In the top panel Raman spectra with a polarization angle of {$\phi=150^\circ$} are depicted, while the panels on the right show the integrated intensity in the range of the $\mathrm{B^{(5)}_g}$/$\mathrm{A^{(9)}_g}$ pair and the corresponding slice of the fit. The Raman spectra are depicted with the square root of the intensity for better visualization of small peaks.}
\label{fig:PARRS_201}
\end{figure}

\clearpage
\subsection{Differences between model fit and Raman spectra for the a-plane}

\begin{figure}[tbh]
\includegraphics[width=0.7\linewidth]{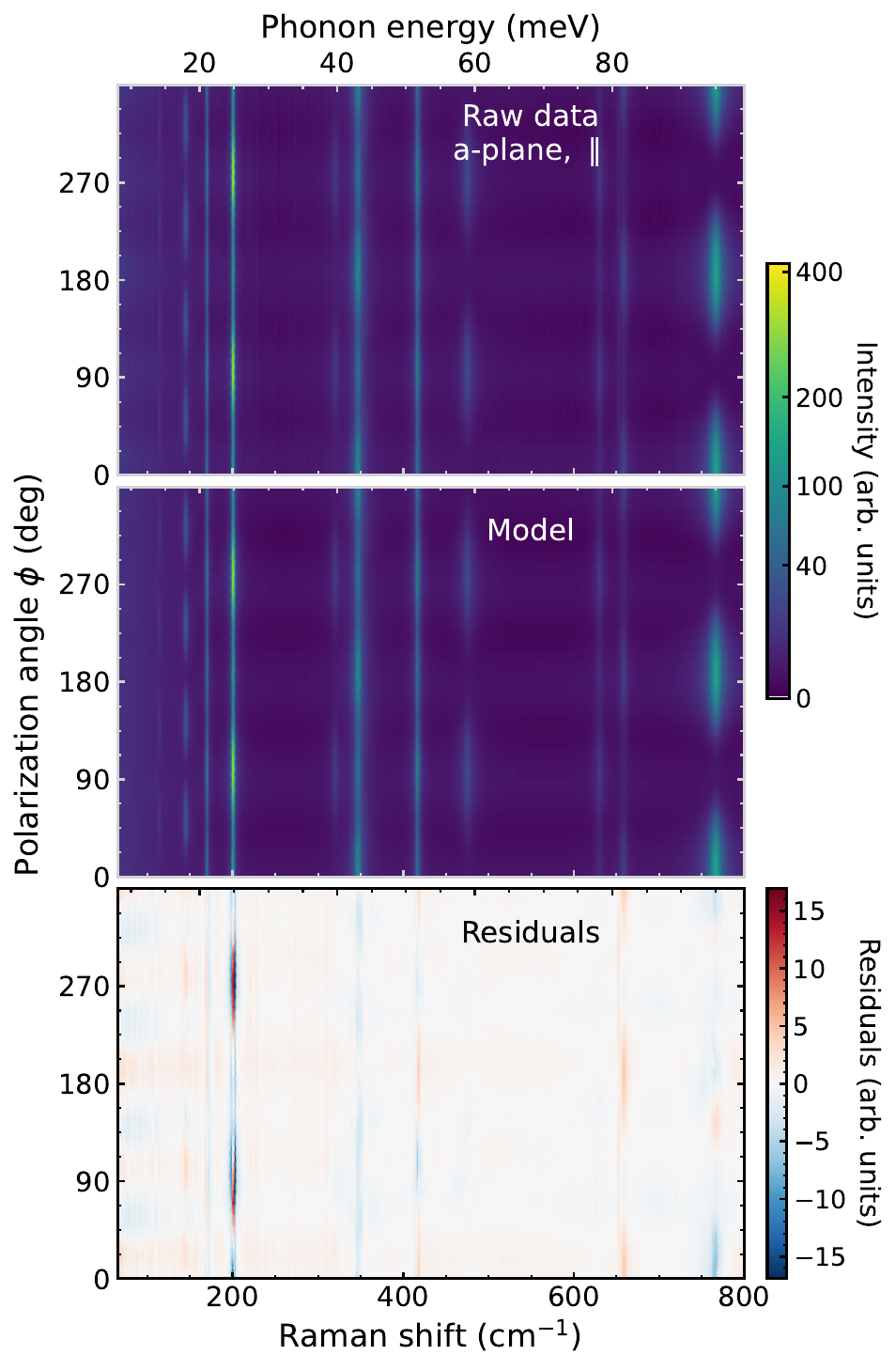}
\caption{Contour plot of the experimental polarization dependent Raman spectra (top), model fit (middle) and their difference (bottom) for the a-plane for parallel scattering geometry.
The color bar for the Raman spectra and model fit is scaled by the square root for better visualization of small peaks.}
\label{fig:residuals_para}
\end{figure}

\begin{figure}[tbh]
\includegraphics[width=0.7\linewidth]{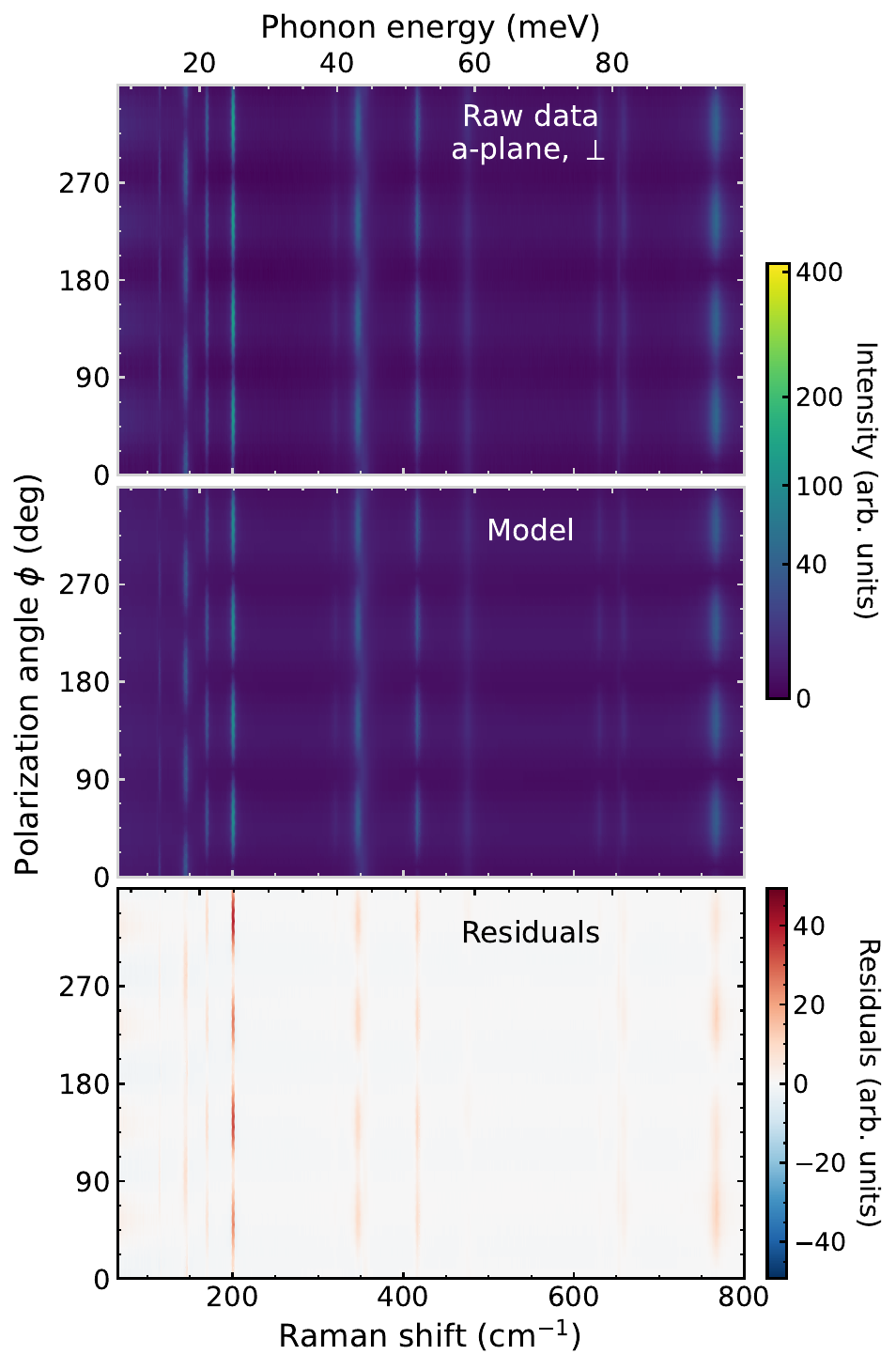}
\caption{Contour plot of the experimental polarization dependent Raman spectra (top), model fit (middle) and their difference (bottom) for the a-plane for crossed scattering geometry.The color bar for the Raman spectra and model fit is scaled by the square root for better visualization of small peaks.}
\label{fig:residuals_perp}
\end{figure}

%%%%%%
% commented text in "*old-text.tex" file
%%%%%%

\end{document}